\documentclass[aoas]{imsart}

\RequirePackage{amsthm,amsmath,amsfonts,amssymb}
\RequirePackage[authoryear]{natbib}
\usepackage{authblk}
\usepackage{amsmath}
\usepackage{amssymb}
\usepackage{tikz}
\usetikzlibrary{arrows}
\usepackage{float} 
\usepackage{booktabs}
\usepackage{algorithm}
\usepackage{algpseudocode}
\usepackage{bbm}
\usepackage{dsfont}
\usepackage{hyperref}

\DeclareMathAlphabet{\mathbbold}{U}{bbold}{m}{n}

\usepackage[paper=a4paper,margin=1in]{geometry}

\usepackage{adjustbox}
\newtheorem{prop}{Proposition}
\newcommand{\bi}{\begin{itemize}}
\newcommand{\ei}{\end{itemize}}

\usepackage[shortlabels]{enumitem}
\usepackage{subcaption}
\DeclareMathOperator*{\argmin}{argmin}
\DeclareMathOperator*{\argmax}{argmax}
\DeclareMathOperator*{\Argmin}{Argmin_{\epsilon}}

\usepackage{tikz}
\newcommand*\circled[1]{\tikz[baseline=(char.base)]{
            \node[shape=circle,draw,inner sep=0.6pt] (char) {#1};}}

\startlocaldefs

\endlocaldefs

\begin{document}

\null\hfill\begin{tabular}[t]{l@{}}
{
\tiny{Distribution Statement A. Approved for public release: distribution is unlimited}
}
\end{tabular}

\begin{frontmatter}
\title{Expected Diverse Utility (EDU): Diverse Bayesian Optimization of Expensive Computer Simulators}
\runtitle{Expected Diverse Utility}

\begin{aug}
\vspace{4mm}
John J. Miller\textsuperscript{1}, Simon Mak\textsuperscript{1}\footnote{Corresponding author, e-mail: sm769@duke.edu}, Benny Sun\textsuperscript{1},\\
\vspace{2mm}
Sai Ranjeet Narayanan\textsuperscript{2}, Suo Yang\textsuperscript{2}, Zongxuan Sun\textsuperscript{2}\\
\vspace{2mm}
Kenneth S. Kim\textsuperscript{3}, Chol-Bum Mike Kweon\textsuperscript{3}\\

\address[A]{Department of Statistical Science, Duke University}

\address[B]{Department of Mechanical Engineering, University of Minnesota}

\address[C]{DEVCOM Army Research Laboratory, Aberdeen Proving Ground}
\end{aug}

\begin{abstract}
The optimization of expensive black-box simulators arises in a myriad of modern scientific and engineering applications. Bayesian optimization provides an appealing solution, by leveraging a fitted surrogate model to guide the selection of subsequent simulator evaluations. In practice, however, the objective is often not to obtain a single good solution, but rather a ``basket'' of good solutions from which users can choose for downstream decision-making. This need arises in our motivating application for real-time control of internal combustion engines for flight propulsion, where a diverse set of control strategies is essential for stable flight control. There has been little work on this front for Bayesian optimization. We thus propose a new Expected Diverse Utility (EDU) method that searches for diverse ``$\epsilon$-optimal'' solutions: locally-optimal solutions within a tolerance level $\epsilon > 0$ from a global optimum. We show that EDU yields a closed-form acquisition function under a Gaussian process surrogate model, which facilitates efficient sequential queries via automatic differentiation. This closed form further reveals a novel exploration-exploitation-diversity trade-off, which incorporates the desired diversity property within the well-known exploration-exploitation trade-off. We demonstrate the improvement of EDU over existing methods in a suite of numerical experiments, then explore the EDU in two applications on rover trajectory optimization and engine control for flight propulsion.

\end{abstract}


\end{frontmatter}


\section{Introduction}
\label{sec:intro}
With recent fundamental developments in computing algorithms and architecture, scientific computing plays an increasingly essential role in tackling modern scientific and engineering problems. Such progress enables the reliable simulation of highly complex phenomena via virtual experimentation, including universe expansions \citep{kaufman2011efficient}, particle collisions \citep{ji2023graphical,li2023additive} and human organs \citep{chen2021function}. These ``computer experiments'', however, come at a cost: they typically require a considerable amount of computing resources. For example, the virtual simulation of a new rocket engine for space exploration can require millions of CPU hours \citep{yeh2018common}. Furthermore, decision-making with such experiments often involves its optimization over a large parameter space, which requires performing many simulation runs at different parameters \citep{gonzález2015Bayesian}. This introduces a computational bottleneck that greatly hampers the full use of computer experimentation for timely scientific discovery and decision-making.

This computational bottleneck is a critical limitation in our motivating application on real-time control of compression ignition engines for flight propulsion \citep{amezcua2020optical,amezcua2022ignition}. These engines are increasingly used in flight applications, particularly in advancing the frontiers of sustainable aviation for commercial and industrial settings \citep{stafford2023combined}. Here, real-time engine control is crucial for stable flight performance under broad operating conditions, particularly given the increased proclivity of such engines for ignition misfires with sustainable aviation fuels \citep{miganakallu2022impact}. Control strategies are typically optimized via a ``digital twin'' representation of the engine, which can virtually simulate combustion performance at different control parameters. However, \textit{each} run of this virtual simulation requires over 3,000 CPU hours to reliably capture the complex underlying computational fluid dynamics! A naive approach for control optimization (e.g., via line-search methods; see \citealp{nocedal1999numerical}) may require running thousands of simulation runs over the parameter space, which is clearly prohibitively expensive. A more cost-efficient optimization approach is thus desired in practical scenarios where only limited experimental runs can be afforded; more on this in Section \ref{sec:mot}. 


A promising solution for the optimization of expensive black-box simulators is Bayesian optimization (BO; \citealp{frazier2018tutorial}), which tackles the following problem:
\begin{equation}
\mathbf{x}^* = \argmin_{\mathbf{x}} f(\mathbf{x}).
\label{eq:bo}
\end{equation}
Here, $f(\mathbf{x})$ is the simulated output at input parameters $\mathbf{x}$, which lies on a normalized parameter space $\mathcal{X} = [0,1]^d$. In our flight control application, $f$ may be a simulated metric for flight instability and $\mathbf{x}$ may be its control parameters. BO begins with a probabilistic predictive model (or ``surrogate model''; \citealp{gramacy2020surrogates}) on $f(\cdot)$, which is trained using a limited number of simulation runs over the parameter space. Using this trained surrogate, an acquisition function is then defined for selecting the next point $\mathbf{x}_{\rm new}$ for subsequent evaluation of $f$. For surrogate modeling, the Gaussian process (GP; \citealp{RasmussenW06}) is widely used due to its closed-form posterior predictive equations. For the acquisition function, a popular choice is the Expected Improvement (EI) acquisition \citep{Jones1998}. A key reason for its popularity is its closed-form expression under a GP surrogate \citep{chen2023}, which permits effective optimization of subsequent evaluation points via automatic differentiation \citep{baydin2018automatic}; more on this later. Another choice of acquisition is the knowledge gradient \citep{knowledgegradient2008}, which uses a one-step look-ahead policy for improving the optimum of the posterior mean function on $f$. Its acquisition, however, requires costly Monte Carlo approximations; \cite{gramacy2022triangulation} proposed a potential workaround via a careful choice of Delaunay triangulation candidates. Recent developments on BO investigate various robust extensions for real-world decision-making, including adversarially robust BO \citep{Bogunovic2018,christianson2023robust} and robust BO under noisy inputs \citep{Williams2000,miller2024targeted}.

For our flight control problem, however, a further \textit{diversity} requirement is needed for Bayesian optimization. To achieve stable flight, it is desirable to have a basket of multiple ``diverse'' optimal control solutions spread out over the parameter space; such a need arises from a number of practical restrictions. First, the CI engines rely on precise blends of sustainable aviation fuels with standard jet fuels \citep{stafford2023combined, miganakallu2022impact, amezcua2022ignition}. However, during flight, there are considerable restrictions on available fuel blends; such restrictions can further vary during operation due to changing flight conditions and/or weight limitations. Second, one also encounters restrictions on engine energy consumption, which may again vary during operation due to changing environmental and/or flight conditions. Having a diverse basket of optimal control strategies permits not only the timely \textit{accommodation} of such varying constraints in real-time, but also the \textit{planning} of flight strategies to ensure fuel efficiency. Despite the vast body of work on BO, there is scant literature on integrating diversity for black-box optimization. A recent work \citep{maus2023discovering} proposes a promising approach called ROBOT, which tackles diverse Bayesian optimization via a hierarchical rank-ordering of a collection of trust regions. However, as shown later in our experiments, ROBOT may experience mediocre optimization performance given the highly limited sample sizes from expensive simulators.


We thus propose a new Expected Diverse Utility (EDU) method, which tackles this diverse optimization problem for expensive black-box simulators. While the EDU targets our aforementioned engine control problem, it has broad use for a myriad of applications where diverse optimization is desirable, e.g., in material science, path planning and policy-making; more on this later. The EDU aims to pinpoint diverse ``$\epsilon$-optimal'' solutions: locally-optimal solutions within a tolerance level $\epsilon>0$ from a global optimum. Similar to EI, the EDU provides a closed-form acquisition under a GP surrogate model, which allows for efficient sequential queries using automatic differentiation methods \citep{baydin2018automatic}. This closed form further reveals a novel \textit{exploration-exploitation-diversity} trade-off, which incorporates the desired diversity behavior within the well-known exploration-exploitation trade-off from reinforcement learning \citep{kearns2002near}. We illustrate the effectiveness of EDU over existing methods in a suite of numerical experiments and in two applications: the first on rover trajectory optimization and the second on our motivating aviation engine control problem.


The paper is organized as follows. Section \ref{sec:mot} provides background on the motivating flight control application and its need for diverse black-box optimization. Section \ref{sec:dei} outlines the proposed EDU approach, including the derivation of its closed-form acquisition function and an examination of its properties. Section \ref{sec:method} explores methodological developments on the EDU, including batch sampling and acquisition optimization. Section \ref{sec:num_exper} compares EDU to the state-of-the-art in a suite of numerical experiments. Section \ref{sec:app} demonstrates the effectiveness of EDU in the above two applications. Section \ref{sec:conc} concludes the paper.

\section{Real-Time Engine Control for Sustainable Aviation}
\label{sec:mot}

We first outline our motivating aviation engine control application, and then highlight its need for diverse black-box optimization.

\subsection{Problem Background}
\label{sec:prob}
Internal combustion engines are power-generating devices in which fuel is ignited to generate energy; these engines are broadly used in commercial vehicles and aircraft. In recent years, there has been much promising work on the use of one such engine, the compression ignition (CI) engine \citep{amezcua2020optical,amezcua2022ignition,stafford2023combined}, for sustainable aviation applications. One appeal of CI engines is that they run on blends of sustainable aviation fuel (SAF), which are cleaner and more environmentally friendly than conventional fuels. This, however, comes at a cost: sustainable fuels may encounter difficulties with ignition compared to traditional fossil fuels, e.g., diesel. In particular, for flight applications, the ignition of certain SAF blends becomes challenging at high altitude conditions of low temperature and pressure \citep{stafford2023combined,miganakallu2022impact,amezcua2022ignition}. Such ignition failures may arise due to insufficient in-cylinder temperatures, resulting in so-called ignition ``misfires'' that lead to severe engine damages. To address this, one requires an ignition-assisting mechanism called a glow plug, which is a high temperature rod located near the engine fuel injector to trigger fuel ignition. Further background on the design and use of such engines for sustainable aviation can be found in \cite{sapra2023numerical,sapra2024computational,kumar2024numerical,nejadmalayeri2023multi,narayanan2024misfire,narayanan2023physics}.

\begin{figure*}[!t]
\begin{center}
    \centering
    \includegraphics[width=0.50\linewidth]{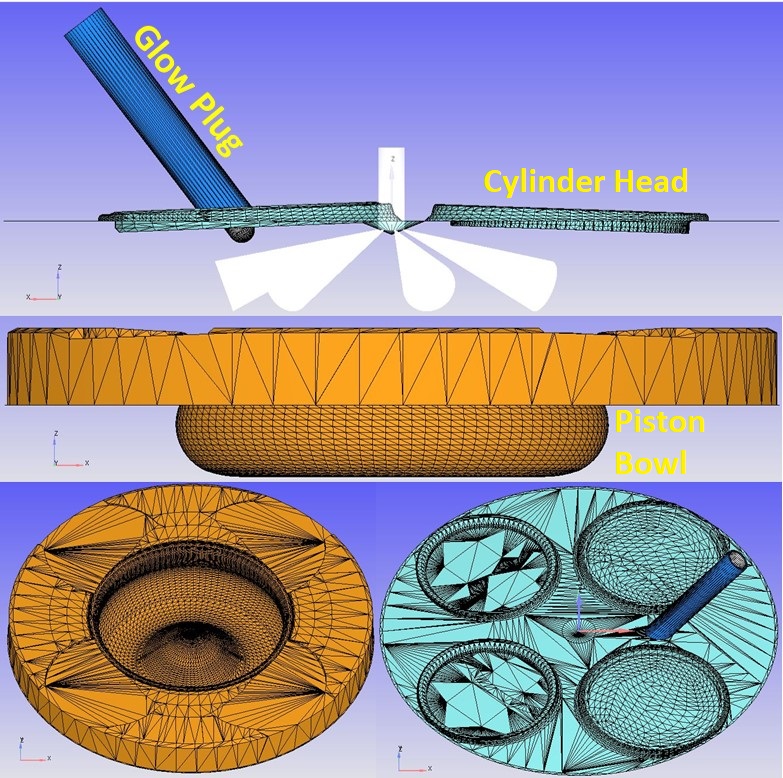}
    \caption{Schematic of the considered metal engine for CFD simulation. Here, the piston bowl and cylinder head are shown, and fuel injection occurs via seven nozzles (marked by white cones). The glow plug (in blue) is seen attached to the cylinder head.}
    \label{geom1}
\end{center}        
\end{figure*}

Figure~\ref{geom1} shows a schematic of the specific CI engine studied in our work. This engine is a General Motors single-cylinder CI metal engine fitted with a glow plug ignition assistant. Such an engine has been explored in physical experiments for multi-fuel injection (i.e., blends of SAFs and jet fuels) at varying flight conditions \citep{stafford2023combined,miganakallu2022impact,amezcua2022ignition} and control system settings \citep{dong2022data,govindrajurate2023,pal2024data}. To explore performance over a broad range of control parameters, we make use of a ``digital twin'' representation (see, e.g., \citealp{jones2020characterising}) of this engine, which virtually simulates its underlying complex fine-scale combustion behavior. This digital twin is built via the Converge CFD v3.0 software \citep{richards2020converge}, and captures the desired engine geometry (Figure \ref{geom1}) within the simulator model. In our experiments, simulation runs were performed on a high-performance computing (HPC) node containing 128 AMD 3rd-Gen EPYC Milan processors with 512 GB RAM. Despite the use of HPC, each simulation run is highly expensive computationally; each run takes, on average, around 24 hours to perform (or equivalently, around 3,000 CPU hours). One reason for this cost is the need for modeling the fine-scale multi-physics nature of complex combustion mechanisms, including turbulence flows, fuel combustion, detailed chemical reactions, and species/fluid transport. Table~\ref{table_engine_models} (summarized from \citealp{sapra2023numerical}) outlines the employed simulation framework.



\begin{table*}[!t]
\centering
\begin{minipage}{.64\linewidth}
\centering
\caption{\label{table_engine_models} Simulation set-up for the considered CI engine digital twin. }
\resizebox{\linewidth}{!}{
\begin{tabular}{cc}
\hline
\hline
\textit{Modeled Physics} & \textit{Simulation Approach}\\
\hline
Combustion & SAGE Chemical Kinetics Solver \citep{senecal2003multi}\\
Chemical Mechanisms & Multicomponent (178 species, 758 reactions) \citep{ren2017multi}\\
Turbulence Flow & RANS model (RNG $k-\epsilon$) \citep{tsan1995new}\\
Droplet Evaporation & Frossling model \citep{amsden1989computer}\\
Spray Breakup	& KH-RT model \citep{reitz1986mechanism}\\
Collision	& No-Time-Counter collision model \citep{schmidt2000new}\\
\hline
\hline
\end{tabular}
}
\end{minipage}%
\hfill
\begin{minipage}{.35\linewidth}
\centering
\caption{\label{table_range} Control parameter ranges for the considered CI engine.}

\begin{tabular}{cc}
\hline
\hline
\textit{Parameter} & \textit{Range} (units)\\
\hline
$x_1$: SOI & -25 -- 0 (CAD)\\
$x_2$: GPP & 0 -- 70 (Watts)\\
$x_3$: CN & 25 -- 48\\
$x_4$: RPM & 1200 -- 2400 (RPM)\\
\hline
\hline
\end{tabular}

\end{minipage}
\end{table*}


Given the critical risk for ignition misfires, timely engine control is thus imperative for stable flight performance \citep{sun2014design}. Such control requires a careful optimization of control parameters, such as fuel injection timing and glow plug power, over different combinations of fuel blends, engine speeds and ambient conditions. To achieve \textit{real-time} control, existing methods employ feed-forward lookup tables \citep{pal2024data,govindrajurate2023,narayanan2024simulation}. These lookup tables are generated in an offline fashion (i.e., prior to flight), by first performing a batch of simulation runs uniformly over the control parameter space. During flight, the controller then ``looks up'' a simulated control setting that optimizes a desired performance metric given current operating conditions. Lookup tables, however, have a key limitation: the computational cost needed to saturate the control parameter space with simulation runs is prohibitively high, given the expensive nature of each run. This greatly restricts the use of traditional look-up tables for full-scale sustainable aviation applications \citep{narayanan2024misfire}. Bayesian optimization offers a promising solution: it leverages a fitted surrogate model to guide an \textit{informed} sampling of the parameter space, thus facilitating improved optimization performance (and better flight control) with fewer simulation runs.


In our study, we investigate $d=4$ control parameters for engine control: the start-of-injection time (SOI; denoted as $x_1$), glow plug power (GPP; denoted as $x_2$), engine speed (measured in revolutions-per-minute or RPM; denoted as $x_3$) and fuel cetane number (CN, denoted as $x_4$). The fuel CN is an indicator for fuel ignitability: a higher CN fuel ignites more easily (i.e., earlier) compared to a lower CN fuel. Such ignitability can be varied by changing different blends of fuel. The current CI engine follows a single four-stroke cycle (air intake, compression, expansion, emission discharge), ranging from -360 to +360 crank-angle-degrees (CADs). Following the literature on CI engines \citep{sapra2023numerical,narayanan2024misfire}, we use CAD as a measure of time scale here. The SOI control parameter $x_1$ is then defined on $[-25,0]$ CAD. Table \ref{table_range} summarizes the considered ranges for control parameters in our case study.




\subsection{A Need for Diverse Solutions}
\label{sec:need}



To achieve stable real-time flight control, it is desirable to have a \textit{diverse} basket of optimal control solutions spread over the parameter space. This need arises largely from changing constraints on {fuel blends} and engine {energy consumption} during flight, which in turn imposes changing restrictions on the $d=4$ control parameters. For fuel blends (i.e., of sustainable aviation fuels with conventional jet fuels), a controller will likely encounter shortages for certain types of fuel at different stages of flight \citep{pal2024data,dong2022data,govindrajurate2023}, which restrict the range of blends available for combustion. This is further compounded by the limited storage space on aircrafts from weight limitations, which can restrict the variety of fuel blends carried for operation. For energy consumption, there are critical safety concerns that restrict certain control strategies in different environmental and/or flight conditions. One such condition is at high flight altitudes with low ambient temperature; here, the use of low-CN fuels (i.e., low ignitability) with low glow plug power needs to be avoided for control, as this greatly increases the risk of misfires and subsequent severe engine damage. Another condition is during periods of high energy expenditure; here, the use of high glow plug power needs to be avoided for control, as this accelerates engine breakdown and failure \citep{miganakallu2022impact}. Such restrictions on control strategies can further vary considerably over the course of a flight. With a {diverse} basket of control strategies on hand, a controller can adeptly \textit{accommodate} such varying constraints in {real-time}, and \textit{plan} for appropriate flight strategies to ensure fuel efficiency and safety.

\begin{figure}
    \centering
    \includegraphics[width=1\linewidth]{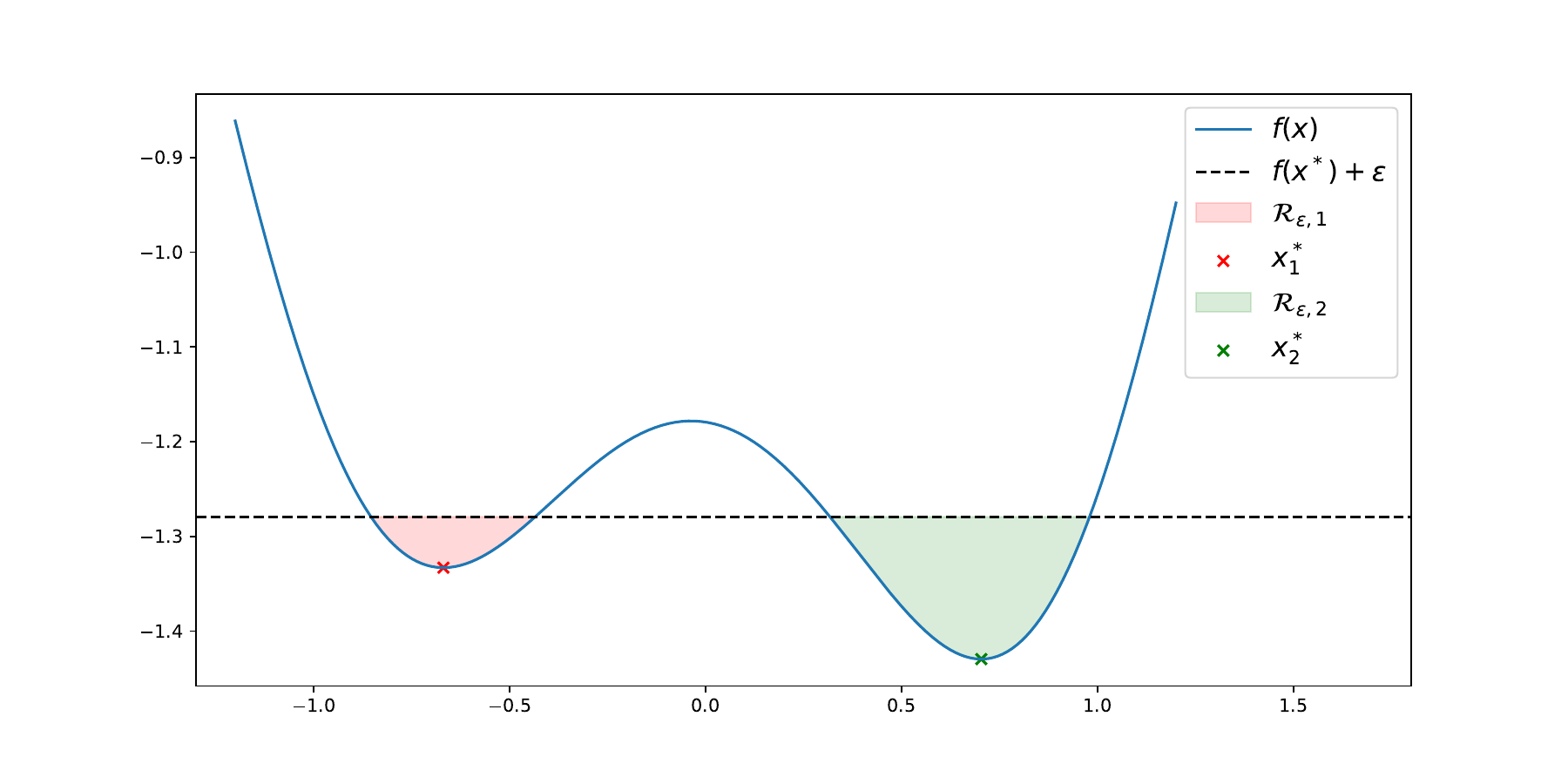}
    \caption{Visualizing a function with $K=2$ diverse optima. Here, with $\epsilon = 0.15$, there are two resulting subregions $\mathcal{R}_{\epsilon,1}$ and $\mathcal{R}_{\epsilon,2}$, with its corresponding minimizers $\mathbf{x}^*_1$ and $\mathbf{x}^*_2$.}
    \label{fig:k-partition}
\end{figure}

A natural formulation of solution diversity is as follows. Let $f(\mathbf{x})$ be a simulated metric for flight instability at control inputs $\mathbf{x} \in \mathcal{X} = [0,1]^d$. Given a pre-specified tolerance level $\epsilon > 0$, a ``tolerable'' solution $\mathbf{x}$ is defined as one for which $f(\mathbf{x}) \leq f(\mathbf{x}^*) + \epsilon$, where $\mathbf{x}^*$ is the global minimizer in \eqref{eq:bo}. In other words, such a solution is within a tolerable amount of $\epsilon$ from the optimal objective $f(\mathbf{x}^*)$. This notion of a tolerance level $\epsilon$ is widely used within standard optimization solvers, and can often be elicited from the domain problem; more on this in Section \ref{sec:eps}. Next, define the ``$\epsilon$-optimal'' region of these solutions as $\mathcal{R}_{\epsilon} = \{\mathbf{x} \in \mathcal{X}: f(\mathbf{x}) \leq f(\mathbf{x}^*) + \epsilon\}$. Here, $\mathcal{R}_{\epsilon}$ may consist of disjoint parts (see Figure \ref{fig:k-partition}); it can thus be partitioned into $K \geq 1$ separate subregions, which we denote as $\mathcal{R}_{\epsilon,1}, \cdots, \mathcal{R}_{\epsilon,K}$. Within each subregion $\mathcal{R}_{\epsilon,k}$, we then wish to find its minimizer $\mathbf{x}^*_k$; this is denoted in the following as:
\begin{equation}
\{\mathbf{x}^*_1, \cdots, \mathbf{x}_K^*\} := \Argmin_{\mathbf{x}} f(\mathbf{x}).
\label{eq:divform}
\end{equation}

\noindent Figure \ref{fig:k-partition} visualizes this diverse optimization formulation, including the partitioning of $\mathcal{R}_\epsilon$ into $K=2$ subregions $\mathcal{R}_{\epsilon,1}$ and $\mathcal{R}_{\epsilon,2}$, with its corresponding local minimizers $\mathbf{x}^*_1$ and $\mathbf{x}_2^*$.

Here, the solutions $\mathbf{x}^*_1, \cdots, \mathbf{x}_K^*$ capture the desired diversity behavior for flight control in an important way. Note that implicit within \eqref{eq:divform} is the prior assumption that, given a pre-specified tolerance level $\epsilon$, the response surface $f$ indeed admits multiple diverse solutions (i.e., $K>1$) over the parameter space. Such diversity is known to be present in our flight application, as existing engine research suggests that different control strategies can yield similarly stable and efficient engine performance \citep{dong2022data,govindrajurate2023,pal2024data}. For each separate subregion $\mathcal{R}_{\epsilon,k}$, we then target an optimal solution $\mathbf{x}_k^*$ localized within such a region. Given a tolerance level $\epsilon$, the basket of solutions $\{\mathbf{x}_1^*, \cdots, \mathbf{x}_K^*\}$ gives optimal control strategies within each tolerable region, which in turn provides flexible control options given changing restrictions in real-time.


It is worth distinguishing the diverse optimization problem \eqref{eq:divform} with the existing work on robust Bayesian optimization surveyed in Section \ref{sec:intro}. The latter aims to minimize the simulated output $f(\mathbf{x},\boldsymbol{\theta})$, where $\mathbf{x}$ and $\boldsymbol{\theta}$ are control and noise (i.e., {uncontrollable}) parameters in reality. Recent works have investigated different forms of uncertainties on $\boldsymbol{\theta}$, including adversarial perturbations \citep{Bogunovic2018,christianson2023robust} and random noise \citep{Williams2000,miller2024targeted}. Compared to this literature, the key distinction in our formulation \eqref{eq:divform} is that \textit{all} parameters are controllable, but there is a myriad of varying restrictions on control parameters that arise during operation. Having a diverse set of solutions on hand allows one to adapt to these restrictions in real-time and plan flight strategies accordingly; such adaptivity is not explored in the above literature.

Our formulation \eqref{eq:divform} is also related to, but distinct from, the problem of contour estimation for black-box functions \citep{ranjancontour}. The latter targets the estimation of the contour region $\{\mathbf{x} \in \mathcal{X}: f(\mathbf{x}) = \gamma\}$ at a given \textit{known} contour level $\gamma$. There is a growing body of work on this for expensive simulators; see, e.g., \cite{ranjancontour,booth2024contour}. Our diverse formulation has several key distinctions with this line of work. First, the identification of the $\epsilon$-optimal region $\mathcal{R}_\epsilon$ in \eqref{eq:divform} involves the contour level $f(\mathbf{x}^*) + \epsilon$, which is \textit{unknown} prior to experimentation as the global minimum $f(\mathbf{x}^*)$ is not known. This contrasts with the contour estimation problem, where the contour level $\gamma$ is known and pre-specified prior to experiments on $f$. Second, the estimation of $\mathcal{R}_\epsilon$ is an intermediate objective in \eqref{eq:divform}; the ultimate goal is to pinpoint \textit{optimal} solutions within its disjoint subregions. This emphasis on optimality is crucial for flight control: while solutions on contour level $f(\mathbf{x}^*) + \epsilon$ may be tolerable temporarily, it is more desirable to identify optimal solutions with lower objective values, particularly when such solutions are known to be present from prior knowledge. In the following section, we will leverage this literature on contour estimation to develop the proposed EDU.


\begin{figure}
    \centering

    \begin{subfigure}{0.45\textwidth}
        \includegraphics[width=\linewidth]{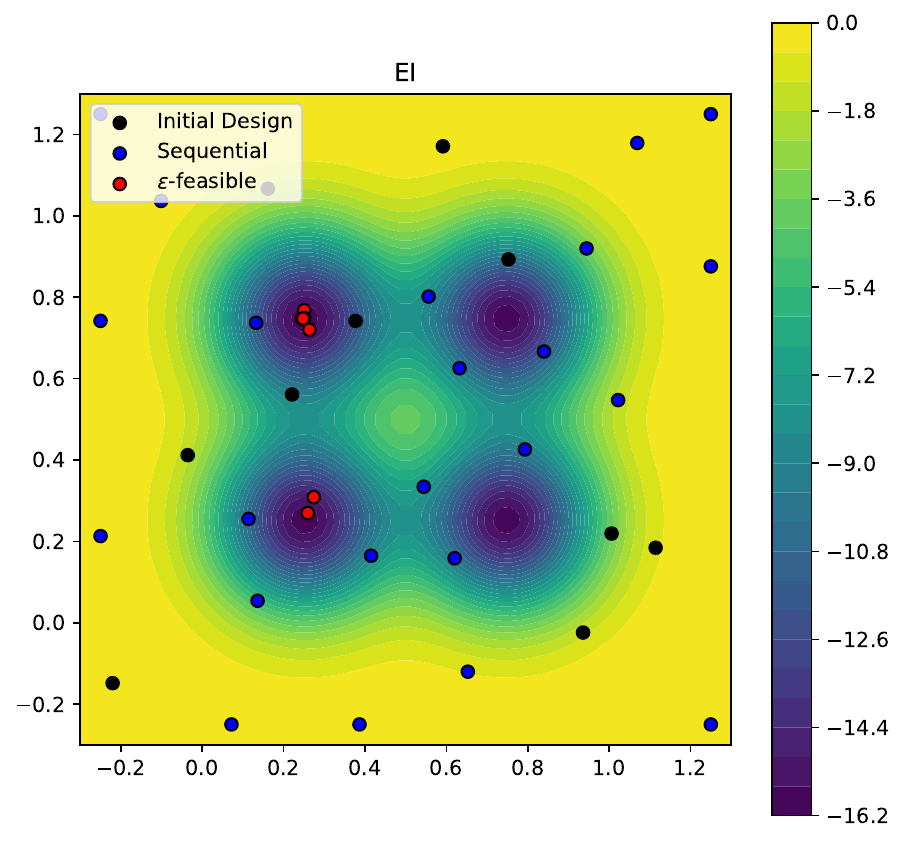}
    \end{subfigure}
    \hfill
    \begin{subfigure}{0.45\textwidth}
        \includegraphics[width=\linewidth]{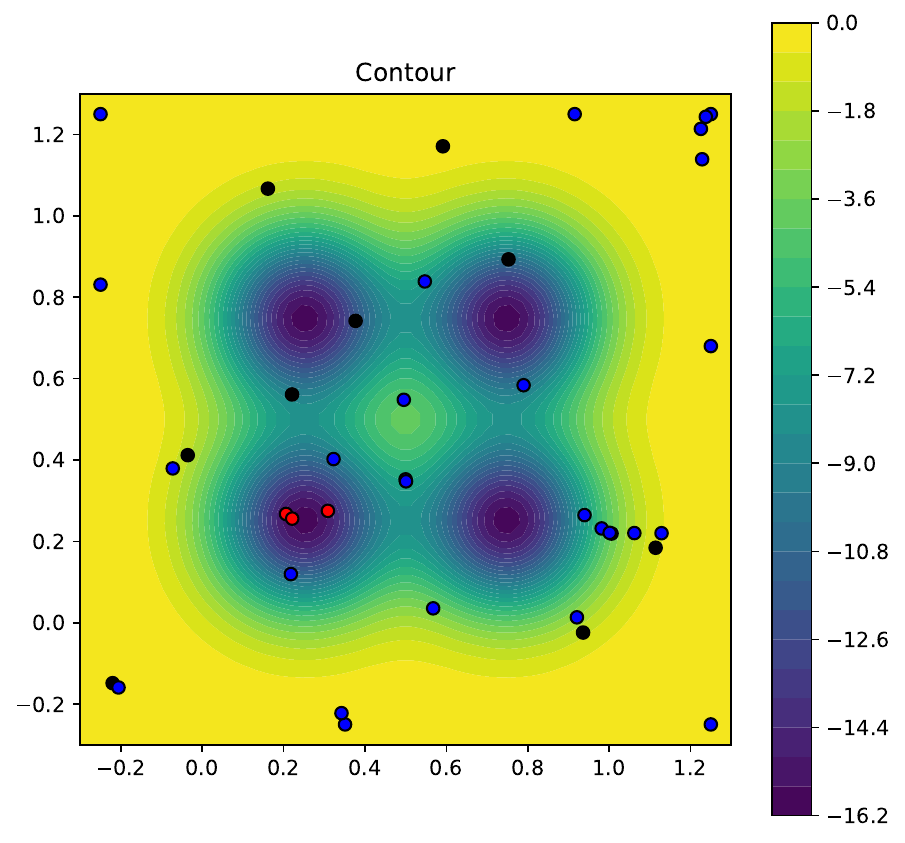}
    \end{subfigure}

    \vspace{1em} 

    \begin{subfigure}{0.45\textwidth}
        \includegraphics[width=\linewidth]{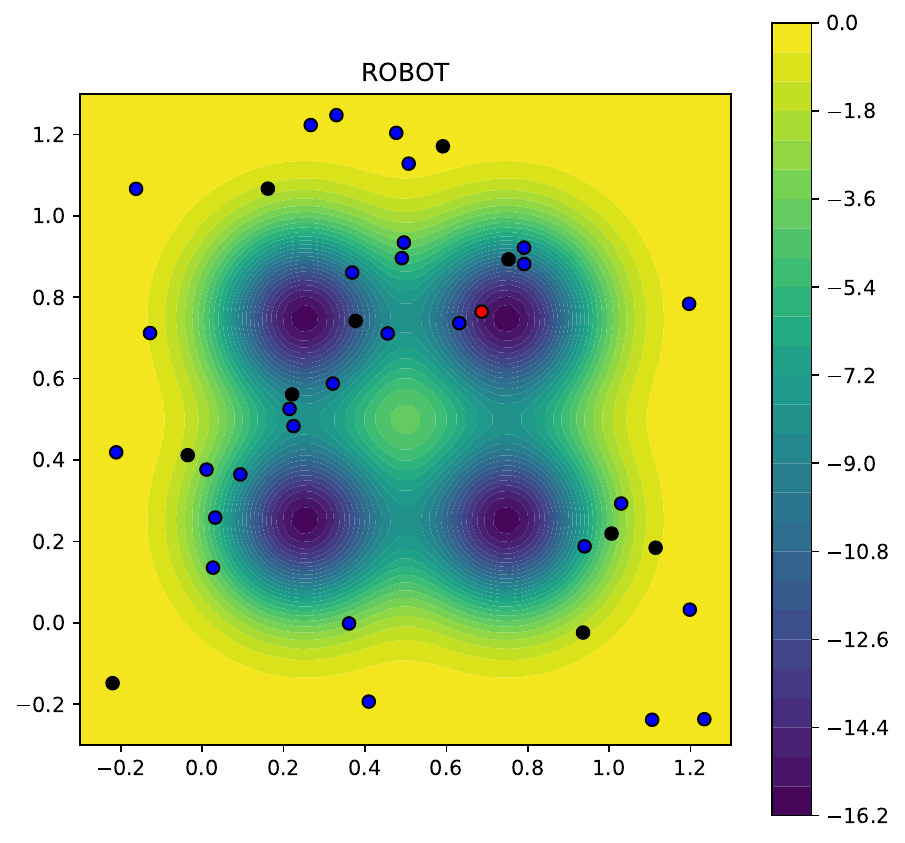}
    \end{subfigure}
    \hfill
    \begin{subfigure}{0.45\textwidth}
        \includegraphics[width=\linewidth]{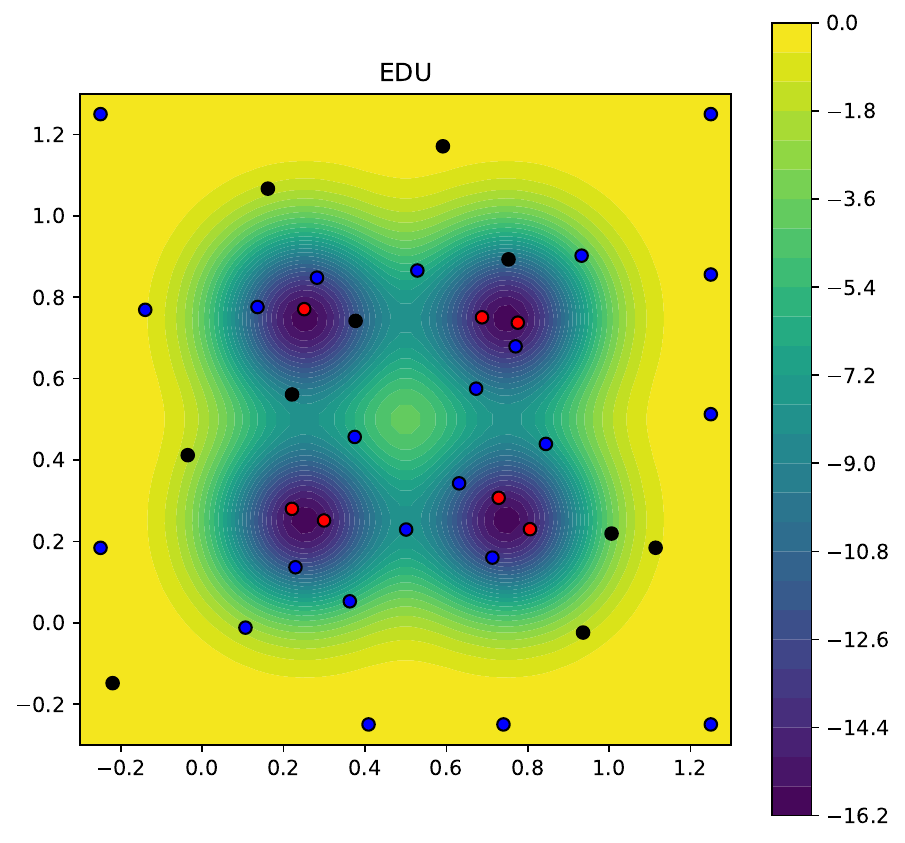}
    \end{subfigure}

    \caption{Visualizing the 2-d four-bowls experiment for diverse optimization. Black points show the $n=10$ initial design points that are shared by all methods. Blue points show the sequential points selected by each method. Red points mark the samples that are within the $\epsilon$-optimal subregion $\mathcal{R}_{\epsilon,k}$ for one of the four optimal solutions.}
    

    \label{fig:4bowls_motivation}
\end{figure}


With this framework in hand, we now investigate key limitations of existing BO methods for the target diverse optimization problem. Consider a simple ``four-bowls'' test function in $d=2$ dimensions, which has $K=4$ global minima at the points $\{0.25,0.75\}^2$. Figure \ref{fig:4bowls_motivation} shows its contour plot, with its full definition provided later in Equation \eqref{eq:bowls}. To achieve diversity, a desirable method should ideally explore all minima within each of its four ``bowls''. We compare here several approaches for selecting sequential points: the EI method \citep{Jones1998}, a standard benchmark for BO; the ROBOT method in \cite{maus2023discovering}, a recent approach for diverse Bayesian optimization via trust regions; and the contour estimation method in \cite{ranjancontour}, where the contour level $\gamma$ is adaptively set as the estimated objective threshold $f_{\rm min} + \epsilon$, and $f_{\rm min}$ is the smallest observed objective from data. All methods use the same initial Latin hypercube design \citep{mckay_LHS} with $n=10$ points, followed by 15 sequential samples. Here, the tolerance level $\epsilon$ is set as $|f(\mathbf{x}^*)|/10$ for diverse optimization.

Figure \ref{fig:4bowls_motivation} shows the sampled design points from each method, with red points marking the samples that are within the $\epsilon$-optimal subregion $\mathcal{R}_{\epsilon,k}$ for one of the four optimal solutions. We see that all methods face notable limitations for diverse optimization. For the EI, its selected points largely focus around the current best solution; as such, it finds only two of the four tolerable optima. This is not surprising, as the EI (and standard BO methods) are designed to target a \textit{single} best solution (see \eqref{eq:bo}) rather than a basket of diverse solutions. In higher dimensions, this fixation on the best solution may lead to deteriorating diverse optimization performance, as we show later. For the contour estimation approach, it finds only one of the four tolerable optima. This poor diverse optimization is expected, as such a method targets function contours but not optimal solutions within contours. For ROBOT, it similarly finds only one of the four tolerable optima. We shall see in later experiments that, given the present bottleneck of limited samples, ROBOT may yield poor diverse optimization performance. To foreshadow, the proposed EDU finds all four tolerable optima given the same limited experiments setting; we present this method next.

\section{Expected Diverse Utility}
\label{sec:dei}

We now outline our Expected Diverse Utility method, which targets the black-box optimization of the diverse formulation \eqref{eq:divform}. We first provide a brief overview of GP surrogates for Bayesian optimization. We then present a new ``diverse utility'' function, and derive its closed-form acquisition function under a GP surrogate on $f$. Finally, we investigate how this acquisition targets diverse optimization solutions in an illustrating numerical example.


\subsection{Gaussian Process Modeling}

In what follows, we adopt the following GP model \citep{RasmussenW06} on the black-box function $f$:
\begin{equation}\label{eq:gp_statement}
f(\cdot) \sim \text{GP}\{\mu, k_{\boldsymbol{\theta}}(\cdot,\cdot)\}.
\end{equation}
Here, $\mu$ is its mean parameter, and $k_{\boldsymbol{\theta}}(\cdot,\cdot)$ is its covariance kernel function with parameters $\boldsymbol{\theta}$. Common kernel choices for computer experiments include the squared-exponential and Mat\'ern kernels \citep{gramacy2020surrogates}; in our later experiments, we use the anisotropic squared-exponential kernel with length-scale parameters $\boldsymbol{\theta} \in \mathbb{R}_+^d$.

Now suppose the simulator $f$ has been evaluated on design points $\mathbf{x}_1, \cdots, \mathbf{x}_n \in \mathcal{X}$, yielding data $\mathcal{D}_n = \{f(\mathbf{x}_1), \cdots, f(\mathbf{x}_n)\}$. For our flight application, the simulator is \textit{deterministic}, in that $f$ is observed without any simulation noise; our approach, however, can be easily extended for simulators with Gaussian noise (see, e.g., \citealp{gramacy2020surrogates}). One appeal of GPs is that, conditional on data $\mathcal{D}_n$, the posterior predictive distribution of $f$ at a new point $\mathbf{x}_{\rm new}$ takes the closed form $f(\mathbf{x}_{\rm new}) | \mathcal{D}_n \sim \mathcal{N}\{\mu_n(\mathbf{x}_{\rm new}), \sigma^2_n(\mathbf{x}_{\rm new})\}$, where:
\begin{equation}
\mu_n(\mathbf{x}_{\rm new}) = \mu + \mathbf{k}_n^\top (\mathbf{x}_{\rm new}) \mathbf{K}_n^{-1}(\mathbf{f}_n - \mu \mathbf{1}), \;\;
\sigma^2_n(\mathbf{x}_{\rm new}) = k(\mathbf{x}_{\rm new},\mathbf{x}_{\rm new}) - \mathbf{k}_n^\top (\mathbf{x}_{\rm new})\mathbf{K}_n^{-1}\mathbf{k}_n(\mathbf{x}_{\rm new}).
\label{eq:gpclosed}
\end{equation}
Here, $\mathbf{f}_n = [f(\mathbf{x}_i)]_{i=1}^n$, $\mathbf{K}_n = [k(\mathbf{x}_i,\mathbf{x}_j)]_{i,j=1}^n$, and $\mathbf{k}_n(\mathbf{x}_{\rm new}) = [k(\mathbf{x}_i,\mathbf{x}_{\rm new})]_{i=1}^n$. In practice, the parameters $\mu$ and $\boldsymbol{\theta}$ can be estimated from data and plugged into the above predictive equation \citep{gramacy2020surrogates}. The closed-form distribution \eqref{eq:gpclosed} facilitates efficient downstream uses of the surrogate model, including uncertainty quantification \citep{mak2018efficient}, inverse problems \citep{everett2021multisystem,everett2021phenomenological} and optimization (\citealp{chen2023}; see next). 

When the goal is to identify a single optimal solution from the black-box function $f$, the popular Expected Improvement method \citep{Jones1998} uses the utility function $\text{I}(\mathbf{x}) = (f_{\rm min} - f(\mathbf{x}))_+ := \max\{f_{\rm min} - f(\mathbf{x}),0\}$, where $f_{\rm min} = \min_{i=1, \cdots, n} f(\mathbf{x}_i)$. Here, $\text{I}(\mathbf{x})$ quantifies the \textit{improvement} of $f(\mathbf{x})$ over the current best solution $f_{\rm min}$. Since $f(\cdot)$ is unknown, we take the expectation of this utility function under the posterior distribution $[f(\cdot)|\mathcal{D}_n]$, yielding:
\begin{equation}
\text{EI}(\mathbf{x}) = \mathbb{E}[\text{I}(\mathbf{x})|\mathcal{D}_n] = \Phi\left( \frac{f_{\rm min}-\mu_n(\mathbf{x})}{\sigma_n(\mathbf{x})} \right) (f_{\rm min} - \mu_n(\mathbf{x})) + \phi\left( \frac{f_{\rm min}-\mu_n(\mathbf{x})}{\sigma_n(\mathbf{x})} \right) \sigma_n(\mathbf{x}).
\label{eq:ei}
\end{equation}
\normalsize
Here, $\Phi$ and $\phi$ are the standard normal c.d.f. and p.d.f., respectively. This provides a \textit{closed-form} acquisition function, from which one can efficiently optimize for a subsequent sample $\mathbf{x}_{\rm new} \leftarrow \text{argmax}_{\mathbf{x} \in \mathcal{X}} \text{EI}(\mathbf{x})$ via automatic differentiation \citep{baydin2018automatic}.

The EI acquisition \eqref{eq:ei} further admits a natural interpretation in terms of the \textit{exploration-exploitation trade-off} (see, e.g., \citealp{chen2023}). To see why, note that the first term in \eqref{eq:ei} increases with $f_{\rm min} - \mu_n(\mathbf{x})$, thus its maximization targets points $\mathbf{x}$ with small fitted objective values $\mu_n(\mathbf{x})$ from the GP surrogate. In other words, this term \textit{exploits} the fitted GP surrogate model for minimizing $f$. Similarly, the second term in \eqref{eq:ei} increases with $\sigma_n(\mathbf{x})$, thus its maximization targets points $\mathbf{x}$ with large uncertainty from the GP surrogate. As such, this term encourages \textit{exploration} of $f$, guided by the posterior uncertainty from the GP. Such an exploration-exploitation trade-off is fundamental for reinforcement learning \citep{kearns2002near}, and is nicely captured within the interpretable closed-form EI acquisition.

\subsection{Diverse Utility}\label{sec:di_utility}

Next, we introduce our new \textit{diverse utility} function $\text{DU}(\mathbf{x})$, which targets the diverse optimization problem \eqref{eq:divform}. We first present this utility function, then justify how each part contributes to the desired goal of diverse optimization. Our utility is given by:
\begin{equation}
\text{DU}(\mathbf{x}) = \begin{cases}
\lambda^2\sigma_n^2(\mathbf{x})+\sigma_n^2(\mathbf{x})(f(\mathbf{x})-\gamma_n)^2,&\text{if } f(\mathbf{x}) < \gamma_n,\\
    \lambda^2\sigma_n^2(\mathbf{x})-(f(\mathbf{x})-\gamma_n)^2,&\text{if } \gamma_n \leq f(\mathbf{x}) \leq \gamma_n + \lambda \sigma_n(\mathbf{x}),\\
    0,& \text{if } f(\mathbf{x}) > \gamma_n+\lambda\sigma_n(\mathbf{x}),
\end{cases}
\label{eq:di}
\end{equation}
where $\gamma_n = f_{\rm min} + \epsilon$ is our current estimate of the objective threshold $\gamma = f(\mathbf{x}^*) + \epsilon$ and $\lambda>0$ is a tuning parameter. While \eqref{eq:di} looks quite involved, its form is motivated by two intuitive goals for diverse optimization: (i) the estimation of the $\epsilon$-optimal region $\mathcal{R}_\epsilon$, and (ii) the identification of optimal solutions within its component subregions $\{\mathcal{R}_{\epsilon,k}\}_{k=1}^K$. We elaborate on each of these goals below.

Consider first goal (i), which targets the \textit{estimation} of the $\epsilon$-optimal region $\mathcal{R}_\epsilon$. In the case where the objective threshold $\gamma = f(\mathbf{x}^*) + \epsilon$ is \textit{known}, this reduces to the contour estimation problem in \cite{ranjancontour}, which targets the estimation of the contour region $\{\mathbf{x} \in \mathcal{X}: f(\mathbf{x}) = \gamma\}$. There, the following contour utility function is proposed:
\begin{equation}
\text{I}_{\rm ctr}(\mathbf{x}) = \begin{cases}
    \lambda^2\sigma_n^2(\mathbf{x})-(f(\mathbf{x})-\gamma)^2,&\text{if } \gamma - \lambda \sigma_n(\mathbf{x}) \leq  f(\mathbf{x}) \leq \gamma + \lambda \sigma_n(\mathbf{x}),\\
    0,& \text{otherwise}.
    \end{cases}
    \label{eq:utilcont}
\end{equation}
The intuition is straight-forward. 
For points $\mathbf{x}$ whose objective $f(\mathbf{x})$ is greater than $\lambda \sigma_n(\mathbf{x})$ from the contour level $\gamma$, $\text{I}_{\rm ctr}(\cdot)$ assigns zero utility as such points are too far from the desired contour. Otherwise, $\text{I}_{\rm ctr}(\cdot)$ assigns the utility $\lambda^2\sigma_n^2(\mathbf{x})-(f(\mathbf{x})-\gamma)^2$, which has two parts. The latter part $-(f(\mathbf{x})-\gamma)^2$ targets points on the contour, by assigning greater utility to points $\mathbf{x}$ for which $f(\mathbf{x})$ is close to $\gamma$. Given this, the first part $\lambda^2\sigma_n^2(\mathbf{x})$ encourages exploration on the desired contour, by assigning greater utility to points $\mathbf{x}$ with higher posterior uncertainty. The parameter $\lambda>0$ controls this trade-off between the targeting of the contour and the exploration of uncertain points. An inspection of the DU utility \eqref{eq:di} shows that its latter two cases closely resemble the contour utility \eqref{eq:utilcont}; the only difference is the replacement of $\gamma$ (the objective threshold, which is \textit{unknown} here) with $\gamma_n$ (its estimate using the current best solution). Thus, such cases in $\text{DU}(\cdot)$ target the first goal of estimating the $\epsilon$-optimal region $\mathcal{R}_{\epsilon}$.

Consider next goal (ii), the \textit{optimization} of solutions within subregions of $\mathcal{R}_{\epsilon}$. The first case of the diverse utility \eqref{eq:di} targets this goal. To see why, note that its condition $f(\mathbf{x}) < \gamma_n$ corresponds to an ``improvement'' scenario, where the objective $f(\mathbf{x})$ dips below the current objective threshold $\gamma_n$. To exploit this improvement, the $\text{DU}(\cdot)$ assigns the utility $\lambda^2\sigma_n^2(\mathbf{x})+\sigma_n^2(\mathbf{x})(f(\mathbf{x})-\gamma_n)^2$. Here, the more important term is $(f(\mathbf{x})-\gamma_n)^2$: this encourages points $\mathbf{x}$ that have \textit{larger} improvement over the current objective threshold $\gamma_n$, in a similar fashion as the EI. Remaining terms in this improvement scenario serve to ensure continuity for $\text{DU}(\cdot)$ and differentiability of the subsequent acquisition function, which is important for efficient acquisition optimization \citep{baydin2018automatic}.

\begin{figure}
    \centering
    \includegraphics[width=0.60\linewidth]{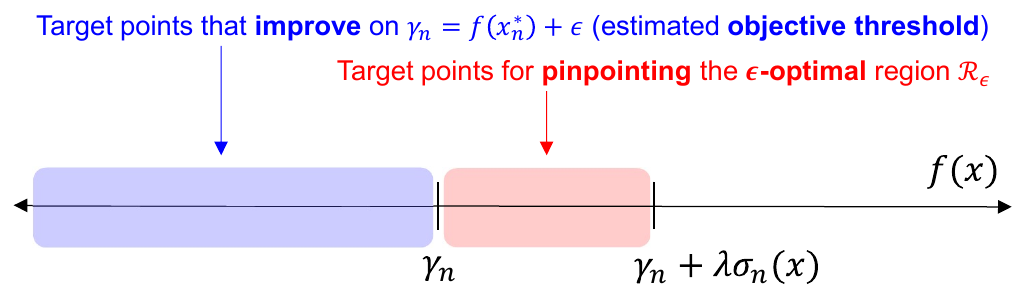}
    \caption{Visualizing the intuition behind the DU utility function in \eqref{eq:di}.}
    \label{fig:DI_intuition}
\end{figure}

Figure \ref{fig:DI_intuition} summarizes the above intuition for the diverse utility function \eqref{eq:di}. The first case in $\text{DU}(\cdot)$, corresponding to $f(\mathbf{x}) < \gamma_n$ and marked in blue, targets the potential for optimization improvement by pursuing points that maximize such improvement (see goal (ii) above). The second case in $\text{DU}(\cdot)$, corresponding to $\gamma_n \leq f(\mathbf{x}) < \gamma_n+\lambda \sigma_n(\mathbf{x})$ and marked in red, targets the estimation of the $\epsilon$-optimal region $\mathcal{R}_\epsilon$ when the point is close to the objective threshold (see goal (i) above). Finally, the third case in $\text{DU}(\cdot)$ assigns zero utility as such points are far from the objective threshold and yield little potential for improvement. Viewed this way, our DU utility can seen as a careful modification of the improvement utility from EI to target the desired goal of diverse optimization.

\subsection{Acquisition Function}
\label{sec:acq}

Next, using the DU utility \eqref{eq:di}, we can then take its posterior expectation under the fitted GP surrogate to derive the so-called Expected Diverse Utility acquisition function. The following proposition shows that this acquisition can be evaluated in closed form:

\begin{prop} Conditional on data $\mathcal{D}_n$, the EDU acquisition function takes the form:
\begin{align}
\begin{split}
    \textup{EDU}(\mathbf{x}) = \mathbb{E}[\textup{DU}(\mathbf{x})|\mathcal{D}_n] &= \left[ \sigma^2_n(\mathbf{x}) + (\gamma_n-\mu_n(\mathbf{x}))^2 \right] \left\{(1+\sigma_n^2(\mathbf{x}))\Phi\left(\zeta_n \right) - \Phi\left(\zeta_n+\lambda \right) \right\}\\ 
    &\quad \quad  + (\gamma_n-\mu_n(\mathbf{x}))\sigma_n(\mathbf{x})\left\{(1+\sigma_n^2(\mathbf{x}))\phi\left(\zeta_n \right) - \phi\left(\zeta_n + \lambda \right) \right\}\\
    & \quad \quad \quad \quad +\lambda \sigma_n^2(\mathbf{x})\left\{\phi\left(\zeta_n+\lambda \right) + \lambda \Phi\left(\zeta_n+\lambda \right)\right\}\\
    &=: \textup{\circled{1} + \circled{2} + \circled{3}},
    \label{eq:dei}
    \end{split}
\end{align}
\noindent where $\zeta_n = (\gamma_n-\mu_n(\mathbf{x}))/\sigma_n(\mathbf{x})$.
\label{prop:dei}
\end{prop}
\noindent While this again looks complicated at first glance, the following natural interpretation can be gleaned upon closer inspection. Consider the first two terms \circled{1} and \circled{2}; we leave \circled{3} for later. Take first the points $\mathbf{x}$ whose predicted mean $\mu_n(\mathbf{x})$ is \textit{close} to the estimated objective threshold $\gamma_n$. Here, the first term \circled{1} clearly dominates \circled{2} via $\sigma^2_n(\mathbf{x}) + (\gamma_n-\mu_n(\mathbf{x}))^2$, and its maximization depends largely on the posterior variance  $\sigma^2_n(\mathbf{x})$. Thus, the EDU targets subregions of the $\epsilon$-optimal region $\mathcal{R}_{\epsilon}$ with high posterior uncertainties, i.e., \textit{diverse} subregions on $\mathcal{R}_{\epsilon}$ that have yet to be explored. Take next the points $\mathbf{x}$ with predicted mean $\mu_n(\mathbf{x})$ considerably \textit{smaller} than the threshold $\gamma_n$. Here, both \circled{1} and \circled{2} grow large via $\sigma^2_n(\mathbf{x}) + (\gamma_n-\mu_n(\mathbf{x}))^2$ and $(\gamma_n-\mu_n(\mathbf{x}))\sigma_n(\mathbf{x})$, respectively, thus the EDU acts aggressively in \textit{exploiting} this improvement from the fitted GP. Finally, take the points $\mathbf{x}$ with $\mu_n(\mathbf{x})$ considerably \textit{larger} than $\gamma_n$, in which case $\zeta_n$ becomes increasingly negative. Note that each term in \circled{1} and \circled{2} shrinks to zero as $\zeta_n$ decreases, thus the EDU does not prioritize sampling within such a region of no improvement.



With this, consider next the third term \circled{3}. Note that the optimization of this term largely depends on $\lambda \sigma_n^2(\mathbf{x})$. By targeting points $\mathbf{x}$ with large $\sigma_n^2(\mathbf{x})$, this term thus targets the \textit{exploration} of regions with high posterior uncertainty from the GP. Putting everything together, the three terms in the EDU acquisition \eqref{eq:dei} nicely capture a novel \textit{exploration-exploitation-diversity} trade-off: they reflect three intuitively desirable properties for diverse black-box optimization, and extend the fundamental exploration-exploitation trade-off for the EI.


This closed form acquisition also offers some insight into the role of the parameter $\lambda > 0$. To see this, consider the following derivative of the EDU acquisition function \eqref{eq:dei} with respect to $\lambda$:
\begin{align}
    \begin{split}
        \frac{\partial \text{EDU}(\mathbf{x})}{\partial \lambda} &= 2\lambda \Phi(\zeta_n+\lambda)\sigma_n^2(\mathbf{x})\\
        & \quad \quad +\phi(\zeta_n+\lambda)\left\{(\zeta_n+\lambda)\left[(\gamma_n-\mu_n(\mathbf{x}))\sigma_n(\mathbf{x})-\lambda \sigma^2_n(\mathbf{x})\right] + (1+\lambda^2)\sigma^2_n(\mathbf{x})\right\}.
    \end{split}
    \label{eq:deriv}
\end{align}
\noindent Note that, when $(\zeta_n+\lambda)$ is positive and large (i.e., the considered point $\mathbf{x}$ is deemed promising for optimization), regions with high uncertainties (i.e., large $\sigma^2_n(\mathbf{x})$) see large partial derivatives for $\partial \text{EDU}(\mathbf{x})/\partial \lambda$ as $\lambda$ increases. A similar phenomenon occurs when $(\zeta_n+\lambda)$ is close to zero. However, when $(\zeta_n+\lambda)$ is large and negative (i.e., the considered point $\mathbf{x}$ is deemed to be suboptimal), this partial derivative becomes closer to zero. Thus, an increase in the parameter $\lambda$ induces a higher preference for \textit{diversity}: it further encourages the search for unexplored solutions only if such solutions are deemed promising for optimization. In our later experiments, we adopt the default choice of $\lambda = 0.5$, which appears to yield a good trade-off for practical diverse optimization problems. 



\begin{figure}[!t]
    \centering

    \begin{subfigure}{0.33\textwidth}
        \includegraphics[width=\linewidth]{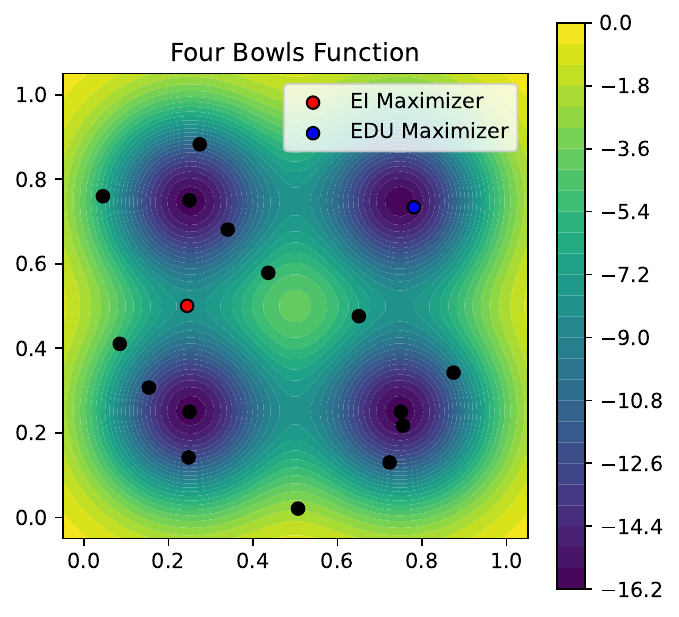}
    \end{subfigure}
    \hfill
    \begin{subfigure}{0.33\textwidth}
        \includegraphics[width=\linewidth]{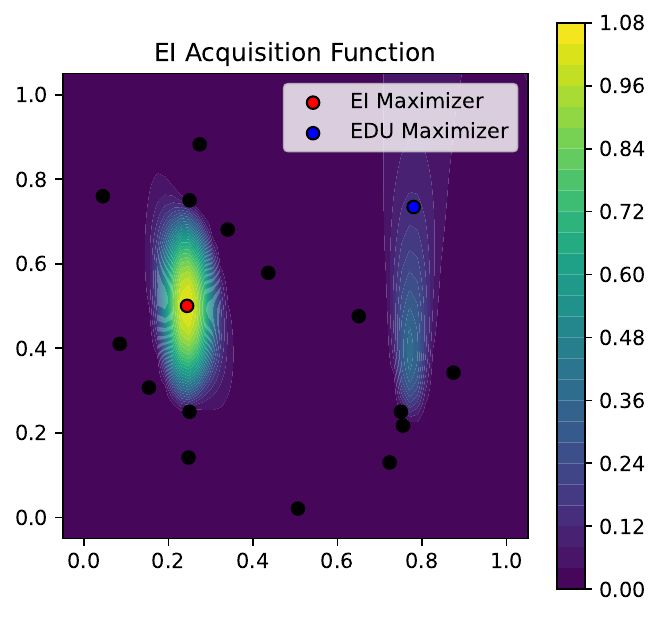}
    \end{subfigure}
    \hfill
    \begin{subfigure}{0.33\textwidth}
        \includegraphics[width=\linewidth]{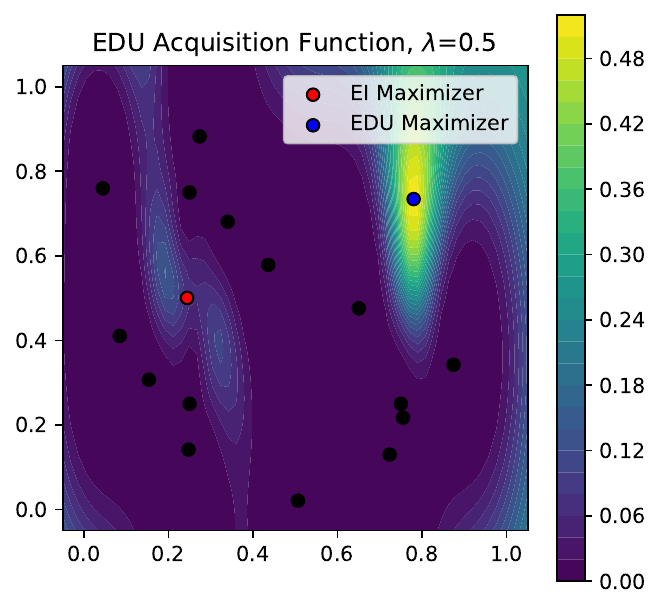}
    \end{subfigure}

    \caption{Comparing the EI and EDU acquisition functions on the four-bowls function (see Section \ref{sec:bowls}) with $n=15$ initial design points. (Left) Initial design points in black, overlaid on function contours of $f$. (Middle and Right) Contours of the EI and EDU acquisitions, respectively, with its corresponding maximizers. 
    }
    \label{fig:EI_DEI_comparison}
\end{figure}

To visualize the improved diversity of the EDU acquisition $\text{EDU}(\mathbf{x})$ compared with $\text{EI}(\mathbf{x})$, we return to the earlier $d=2$-dimensional four-bowls function experiment. In this set-up, $n=15$ initial design points are sampled; such points include three of the four global minima for the function. Figure \ref{fig:EI_DEI_comparison} (left) shows these initial points overlaid on the function contours of $f$; for \textit{diverse} optimization, the goal is to identify the fourth optimal solution in the top-right. Figure \ref{fig:EI_DEI_comparison} (middle and right) shows the contours of the EDU (with $\lambda = 0.5$) and EI acquisitions along with their corresponding maximizers, which will be taken as the next sample points. We see that the EI acquisition focuses heavily within regions near the three observed minima, which is not surprising as the EI targets \textit{any} point that may improve upon the current best solution. The EDU adopts a markedly different strategy. In jointly pursuing the $\epsilon$-optimal region $\mathcal{R}_\epsilon$ as well as its corresponding minima, the EDU places more emphasis on the top-right region, which has greater potential for new $\epsilon$-optimal solutions. This is clearly desirable for diverse optimization here, as the final minimum indeed lies within such a region.

\section{Methodological Developments} 
\label{sec:method}
Next, we discuss methodological developments for practical implementation of the EDU. This includes batch-sequential sampling, acquisition optimization, tolerance specification and determination, and an algorithm statement of the full procedure.

\subsection{Batch EDU}
\label{sec:batch}
In scientific computing, it is often possible (and indeed advantageous) to perform \textit{batch} sequential sampling, where one runs a batch of $q > 1$ simulation experiments simultaneously. This is facilitated by recent breakthroughs in multi-threaded and multi-core processing for high-performance computing \citep{kurzak2010scientific}. As such, the extension of EDU for batch sequential sampling is essential in practical applications; we will tackle this next. 


Again, suppose the simulator $f$ is evaluated at design points $\mathbf{x}_1, \cdots, \mathbf{x}_n$, yielding data $\mathcal{D}_n = \{f(\mathbf{x}_1), \cdots, f(\mathbf{x}_n)\}$. We want to select a next batch of points $\mathbf{x}_{(1)}, \cdots, \mathbf{x}_{(q)}$ for diverse optimization of $f$. There is a body of work on batch sequential sampling for Bayesian optimization. For the EI, \cite{ginsbourger:hal-00260579} proposed the batch acquisition function $\text{I}_{q \text{-EI}}(\mathbf{x}_{(1)}, \cdots, \mathbf{x}_{(q)}) = \max_{j=1, \cdots, q} (f_{\rm min} - f(\mathbf{x}_{(j)}))_+$, which takes the maximum improvement over all $q$ points. The resulting $q$-EI acquisition is then obtained by taking its posterior expectation under a GP surrogate model, and can be obtained via Monte Carlo approximation. A similar batch modification was proposed in \cite{qKG_wu_frazier} for the knowledge gradient. One distinction for the EDU, however, is that we want to encourage \textit{diversity} amongst good solutions; this is not reflected if we took as our utility the maximum improvement over all $q$ points, which targets a single good solution. In particular, here we wish to identify \textit{many} diverse solutions in the sample batch, with such solutions sufficiently different from one another. 




We thus propose the following batch utility function: 
\begin{equation}
q\text{-DU}(\mathbf{x}_{(1)}, \cdots, \mathbf{x}_{(q)}) = \left[1-\max_{j\not=j'}\;\text{Corr}_n\left\{f(\mathbf{x}_{(j)}),f(\mathbf{x}_{(j')})\right\}\right]\sum_{j=1}^q \text{DU}(\mathbf{x}_{(j)}),
\label{eq:qdi}
\end{equation}
\noindent where $\text{Corr}_n\left\{f(\mathbf{x}_{(j)}),f(\mathbf{x}_{(j')})\right\} = \text{Corr}\left\{f(\mathbf{x}_{(j)}),f(\mathbf{x}_{(j')}) | \mathcal{D}_n\right\}$. This captures the desired properties above. The last term in \eqref{eq:qdi} {sums} the diverse utility of point $\mathbf{x}_{(j)}$ over all $j = 1, \cdots, q$, which encourages diverse solutions in the sample batch. The first term in \eqref{eq:qdi} ensures such points are sufficiently far from existing design points and from each other. Note that, in the setting of $q=1$, the $q$-DU utility reduces to the original DU utility in \eqref{eq:di}.

As before, the batch EDU utility \eqref{eq:qdi} naturally yields a closed-form acquisition function via its posterior expectation $q\text{-EDU}(\mathbf{x}_{(1)}, \cdots, \mathbf{x}_{(q)}) = \mathbb{E}[q\text{-DU}(\mathbf{x}_{(1)}, \cdots, \mathbf{x}_{(q)})|\mathcal{D}_n]$. This is given by:
\begin{equation}
q\text{-EDU}(\mathbf{x}_{(1)}, \cdots, \mathbf{x}_{(q)}) = \left[1-\max_{j\not=j'}\;\text{Corr}_n\left\{f(\mathbf{x}_{(j)}),f(\mathbf{x}_{(j')})\right\}\right]\sum_{j=1}^q \text{EDU}(\mathbf{x}_{(j)}).
\label{eq:qdei}
\end{equation}
With this, a new sample batch $\mathbf{x}_{n+1}, \cdots, \mathbf{x}_{n+q}$ can be obtained by maximizing $q\text{-EDU}(\mathbf{x}_{(1)}, \cdots, \mathbf{x}_{(q)})$. The closed form of \eqref{eq:qdei} again facilitates efficient acquisition optimization via automatic differentiation; more on this next.



\subsection{Acquisition Optimization}
\label{sec:opt}
As alluded to previously, a key appeal of the EDU (and in general, EI-based methods; see \citealp{chen2021function}) is that it facilitates efficient acquisition optimization via automatic differentiation. Automatic differentiation \citep{baydin2018automatic} is a rapidly developing area in computer algebra, and has promising use for accelerating gradient-based optimization solvers. The key idea is to automatically compute within a single evaluation of an analytic objective function, e.g., $\text{EDU}(\mathbf{x})$, its corresponding gradient $\nabla_{\mathbf{x}} \text{EDU}(\mathbf{x})$ via the propagation of symbolic differentiation rules. This automatic computation of gradients has two advantages. First, it circumvents the need to analytically derive gradients of complicated functions by hand, thus enabling easy use of gradients with reduced risk for human error. Second, since the gradient $\nabla_{\mathbf{x}} \text{EDU}(\mathbf{x})$ can be directly computed from an objective evaluation of $\text{EDU}(\mathbf{x})$, additional function calls are not needed for evaluating gradients, which considerably speeds up computation. Automatic differentiation has thus seen recent widespread use in many facets of data science, from the training of deep learning models \citep{SCHMIDHUBER201585} to acquisition optimization in reinforcement learning \citep{balandat2020botorch}.

In our implementation, we use the Bayesian optimization framework from the \texttt{BoTorch} library \citep{balandat2020botorch}, in conjunction with the automatic differentiation library in \texttt{PyTorch} \citep{paszke2019pytorch}. For acquisition optimization, we make use of the widely-used L-BFGS-B optimizer \citep{lbgfs-b}. A known challenge with acquisition optimization (see, e.g., \citealp{daulton_ei}) is the non-convex and multi-modal nature of the acquisition function, which makes its global optimization highly challenging. Here, automatic differentiation (with parallel processing) can relax this bottleneck by accelerating multiple restarts of the gradient-based optimizer at different initial points. In our implementation, we found that $4d$-$5d$ restarts of the optimization algorithm yielded good performance, with initial points sampled from a Latin hypercube design \citep{mckay_LHS}.


\subsection{Tolerance Specification and Determination}
\label{sec:eps}
An important practical consideration is the specification of the tolerance level $\epsilon$. As this dictates whether a solution is tolerable or not, such a specification should be directly elicited from the domain problem. Luckily, there is much precedence for error tolerance specification from the standard (i.e., non-black-box) optimization literature; see, e.g., \cite{nocedal1999numerical}. Within state-of-the-art optimization solvers, users typically have the ability to control $\epsilon$, such that the optimized solution satisfies this error tolerance with some degree of confidence. In this setting, practitioners have carefully specified the tolerance level $\epsilon$ from domain knowledge to ensure process quality and reliability; see, e.g., applications in aerospace engineering \citep{bhachu2013aircraft} and lithium ion battery production \citep{batteries6020023}. An analogous user-guided approach can be used for specifying $\epsilon$ in our black-box diverse optimization setting. 

A related question is, after performing the EDU and obtaining evaluated solutions $\{\mathbf{x}_i\}_{i=1}^N$, how do we determine which of these solutions are indeed tolerable? This is deceptively challenging: a solution $\mathbf{x}$ is deemed tolerable if $f(\mathbf{x}) \leq f(\mathbf{x}^*) + \epsilon$, but the global minimum $f(\mathbf{x}^*)$ is typically unknown in practice. One solution is to leverage the practitioner's domain knowledge: given that $\epsilon$ is specified to reflect solution tolerability, the practitioner should be able to identify which of the evaluated solutions are tolerable in implementation. Another solution is to approximate the objective threshold via a known lower bound $f_L$ on the global minimum $f(\mathbf{x}^*)$. For example, in our flight control application, it is known \citep{dong2022data} that the flight instability metric $f(\mathbf{x})$ to minimize is bounded below by $f_L = 0$, and that the global minimum $f(\mathbf{x}^*)$ should be near this lower bound. With $f_L$ identified, an evaluated solution $\mathbf{x}$ can then be deemed tolerable if $f(\mathbf{x}) \leq f_L + \epsilon$; this provides a conservative approach for determining which solutions are tolerable without knowledge of $f(\mathbf{x}^*)$. We employ such a strategy later in our application in Section \ref{sec:ice}.


\subsection{Algorithm Statement}
\label{sec:alg}
We summarize in Algorithm \ref{algorithm:DEI} a full algorithm statement of the EDU procedure for diverse black-box optimization. First, an initial $n_\text{init}$-point design is selected for initial evaluation points $\mathbf{x}_1, \cdots, \mathbf{x}_{n_\text{init}}$. Next, the GP surrogate model (see \eqref{eq:gp_statement}) is trained using the simulated output data. Kernel hyperparameters for the GP are fitted via maximum a posteriori (MAP) estimation \citep{pml1Book}, which is widely used for surrogate training \citep{gramacy2020surrogates}. In our experiments later, this model training is performed via the \texttt{GPyTorch} package \citep{gardner2018gpytorch} in Python. Default priors are used from this package: for the squared-exponential kernel $k$, independent $\text{Gamma}(2,0.15)$ priors are assigned for variance parameters and independent $\text{Gamma}(3,6)$ priors are assigned for length-scale parameters. With the fitted GP in hand, the next evaluation point $\mathbf{x}_{n+1}$ can then be obtained by maximizing $\text{EDU}(\mathbf{x})$; see Section \ref{sec:opt}. Finally, one evaluates the simulator $f(\cdot)$ at this next point, and the procedure is then iterated until the total budget of $N$ evaluations is exhausted. An analogous algorithm holds for the batched EDU (Section \ref{sec:batch}), with the optimization of $\text{EDU}(\cdot)$ replaced by the optimization of $q\text{-EDU}(\cdot)$.

\begin{algorithm}[!t]
    \caption{Expected Diverse Utility for Diverse Black-Box Optimization}
    \raggedright
    \textbf{Input}: Initial design points $\{\mathbf{x}_i\}_{i=1}^{n_{\text{init}}}$, maximum evaluations $N$, diversity parameter $\lambda$.
    \begin{algorithmic}[1]
        \State $\bullet$ Evaluate the expensive simulator $f(\cdot)$ at initial design points $\{\mathbf{x}_i\}_{i=1}^{n_{\text{init}}}$, yielding outputs $\mathbf{f}_{n_{\text{init}}} = [f(\mathbf{x}_i)]_{i=1}^{n_{\text{init}}}$.
        
        \For{$n = n_{\text{init}},\cdots,N-1$}
            \State $\bullet$ Train the GP model  \eqref{eq:gp_statement} using data $\mathbf{f}_n$, with kernel hyperparameters fitted via MAP estimation.
            \State $\bullet$ Optimize the next evaluation point $\mathbf{x}_{n+1} \leftarrow \ \argmax_{\mathbf{x}} \text{EDU}(\mathbf{x})$, where $\text{EDU}(\cdot)$ is as defined in \eqref{eq:dei}.
            \State $\bullet$  Evaluate the expensive simulator $f(\cdot)$ at point $\mathbf{x}_{n+1}$.
            \State $\bullet$ Update the simulated output data $\mathbf{f}_{n+1} = [\mathbf{f}_n; f(\mathbf{x}_{n+1})]$.
        \EndFor
        
   \noindent \hspace{-0.8cm} \textbf{Output}: Evaluated solutions and outputs $\{(\mathbf{x}_i,f(\mathbf{x}_i))\}_{i=1}^N$.
    \end{algorithmic}
    \label{algorithm:DEI}
\end{algorithm}

\section{Numerical Experiments}\label{sec:num_exper}
We now compare the EDU with the existing state-of-the-art in a suite of numerical experiments for diverse optimization. Here, the EDU is implemented with our recommended setting of $\lambda = 0.5$ (see Section \ref{sec:acq}), as well as an alternate setting of $\lambda = 0.25$ for comparison. The following methods are included as benchmarks:
\begin{itemize}
    \item \textit{Expected Improvement} \citep{Jones1998}: The EI serves as our benchmark for standard BO methods. Here, EI is implemented using the \texttt{BoTorch} library \citep{balandat2020botorch} in Python. For fair comparison, the same GP surrogate as the EDU is used, i.e., with a squared-exponential kernel, and kernel parameters estimated via MAP prior to sampling a subsequent point (or batch of points). The same number of random restarts is used for acquisition optimization as the EDU.
    \item \textit{Rank-Ordered Bayesian Optimization with Trust regions} (ROBOT; \citealp{maus2023discovering}): ROBOT is a recent method that tackles the problem of diverse black-box optimization, using a hierarchical rank-ordering of a collection of trust regions. In our experiments, we used the authors' implementation on Github\footnote{\url{https://github.com/nataliemaus/robot}}, with the Euclidean diversity metric $\delta(\mathbf{x},\mathbf{x}') = \| \mathbf{x} - \mathbf{x}'\|_2$. Here, ROBOT is given the advantage of knowing how many $\epsilon$-optimal solutions are present in the true function, as the number of desired solutions is a required input for this method.
    \item \textit{Contour Estimation} \citep{ranjancontour}: The contour estimation approach (discussed in Section \ref{sec:di_utility}) is included as a related benchmark, although such a method is not intended for diverse black-box optimization. For each sequential step, the target contour value $\gamma$ is adaptively set as the estimated objective threshold $\gamma_n = f_{\rm min} + \epsilon$. The same number of random restarts is used for acquisition optimization as the EDU, with $\lambda$ set as 0.5.
    
    \item \textit{Uniform Random Sampling}: Finally, we include the simple baseline of sampling design points uniformly at random over the parameter space.
\end{itemize}
Here, all methods are replicated for 100 trials to show simulation variability. Within a single replicate, all methods share the same initial design, obtained via Latin hypercube sampling \citep{mckay_LHS}. Initial sample sizes are set as $n_{\rm init} = 10d$ following the rule-of-thumb in \cite{loeppky2009choosing}; the only exception is for $d=2$ where $n_{\rm init}$ is reduced to 10 (as $n_{\rm init}=20$ points would well-saturate the low-dimensional space). Similar settings are used for our batched experiments later.




The choice of metric for assessing diverse optimization performance also deserves careful consideration. A standard metric for black-box optimization is the optimization gap $f_{\rm min}-f(\mathbf{x}^*)$, which measures the gap in the objective function between the current best solution and the global minimum. However, while such a metric captures ``pure'' optimization performance, it does not account for the desired diverse behavior of solutions. We thus compare methods on the so-called \textit{solution coverage rate}, defined as the proportion of $\epsilon$-optimal solutions (out of the total $K$ solutions) found by the optimization algorithm. Here, a solution $\mathbf{x}_k^*$ is deemed ``found'' if there exists an evaluated point within its corresponding subregion $\mathcal{R}_{\epsilon,k}$ (see Section \ref{sec:need}). Higher coverage rates thus suggest better diverse optimization performance. In the following experiments, the tolerance level $\epsilon$ is set at $|f(\mathbf{x}^*)|/10$.



\subsection{$2^d$-Bowls Function}\label{sec:bowls}
We first investigate the so-called $2^d$-bowls test function, which extends the earlier four-bowls function. This function is given by:
\begin{equation}
f(\mathbf{x}) = -\sum_{l=1}^{2^d}\phi_d\left(\frac{\mathbf{x}-\boldsymbol{\mu}_l}{\xi} \right), \quad \xi = 0.15.
\label{eq:bowls}
\end{equation}
Here, $\phi_d$ is the $d$-dimensional standard normal density function, and the mean vectors $\{\boldsymbol{\mu}_l\}_l$ are taken from $\{1/4,3/4 \}^d$. By construction, this function has $K=2^d$ $\epsilon$-optimal solutions. The many $\epsilon$-optimal solutions presents a challenging test case for diverse optimization, particularly with limited samples.

\begin{figure}
    \centering

    \begin{subfigure}{0.45\textwidth}
        \includegraphics[width=\linewidth]{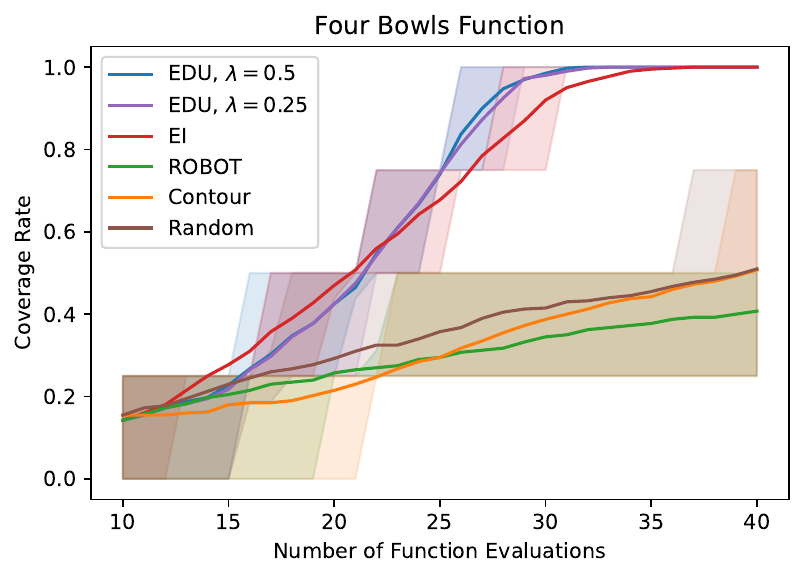}
    \end{subfigure}
    \hfill
    \begin{subfigure}{0.45\textwidth}
        \includegraphics[width=\linewidth]{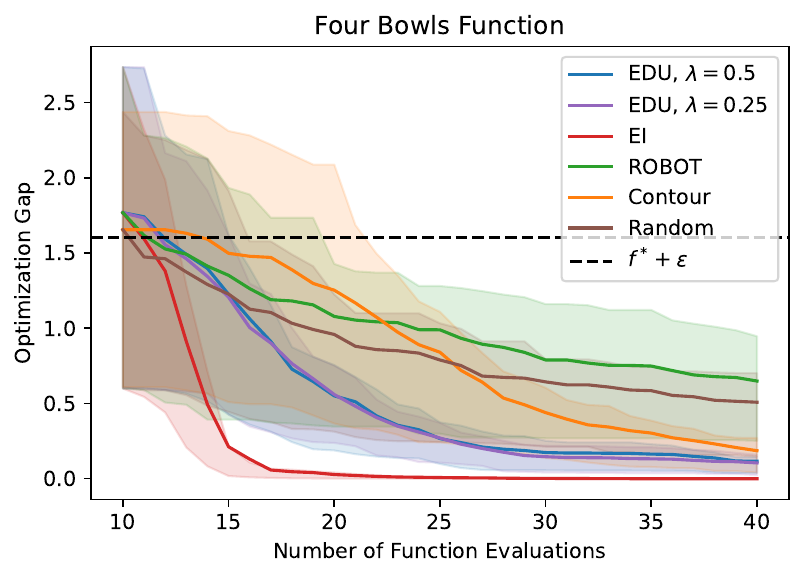}
    \end{subfigure}

    \vspace{1em} 

    \begin{subfigure}{0.45\textwidth}
        \includegraphics[width=\linewidth]{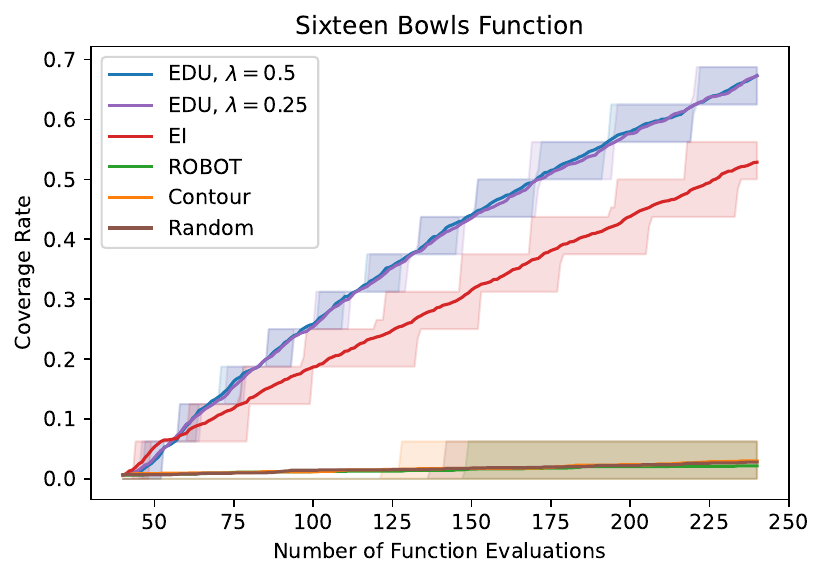}
    \end{subfigure}
    \hfill
    \begin{subfigure}{0.45\textwidth}
        \includegraphics[width=\linewidth]{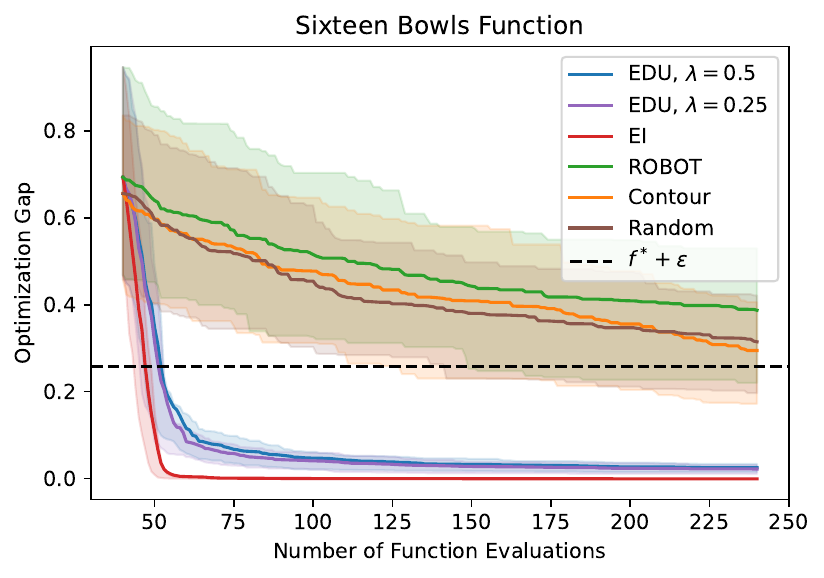}
    \end{subfigure}

    \caption{Solution coverage rates (left) and optimization gaps (right) of the EDU (with $\lambda = 0.5$ and $\lambda = 0.25$), EI, contour estimation, ROBOT and random sampling methods, for the $2^2$-bowl (top) and $2^4$-bowl (bottom) functions. Here, solid lines show the average metric over 100 simulations, with shaded regions showing the 25-th/75-th quantiles. The dotted lines on the right plots show the desired objective threshold $\gamma = f(\mathbf{x}^*) + \epsilon$. 
    }
    \label{fig:bowls_performance}
\end{figure}

Figure \ref{fig:bowls_performance} shows the solution coverage rates and optimization gaps of the compared methods for the $2^2$-bowl and $2^4$-bowl functions, in dimensions $d=2$ and $d=4$, respectively. There are several interesting observations to note. First, the EDU yields considerably higher solution coverage rates over competing methods as more samples are obtained, which indicates better diverse optimization performance. The EI does yield slightly better coverage for the first few sequential samples; this is not surprising upon further inspection, as the EI quickly hones in on the first optimal solution but can struggle in identifying other $\epsilon$-optimal solutions. Second, this improvement in diverse optimization appears to grow as dimension $d$ increases. One reason for this is that, in higher dimensions, the desired $\epsilon$-optimal solutions are more spread apart, and are thus more difficult to pinpoint without careful incorporation of diversity within the acquisition function. Finally, we see that the EDU has a noticeably larger optimization gap compared to the EI. This shows that the desired diverse behavior comes at a cost: by improving solution coverage via searching for $\epsilon$-optimal solutions, the EDU sacrifices some performance for the ``pure'' global optimization problem \eqref{eq:bo}. Such a sacrifice, however, largely happens after the EDU has found solutions satisfying the desired objective threshold $\gamma = f(\mathbf{x}^*) + \epsilon$ (dotted horizontal lines in the right plots in Figure \ref{fig:bowls_performance}).

For the remaining methods, ROBOT yields mediocre performance in terms of both solution coverage and optimization gap; this is not too surprising, as such a method may require a large number of samples to achieve diversity (see \citealp{maus2023discovering}), which cannot be afforded in our setting. In fact, in this limited sample setting, ROBOT is outperformed by random sampling in the $d=2$ experiment. Similarly, the contour estimation method yields poor diverse optimization performance, with coverage rates well below that of the EDU. This again is expected, as such an approach selects points that target contour estimation but not the optimal solutions within such contours. Finally, inspecting the EDU with the alternate setting of $\lambda = 0.25$, we see that such a method yields comparable performance to the default setting $\lambda = 0.5$; we shall see a greater contrast in the higher-dimensional experiments later.



\begin{figure}
    \centering

    \begin{subfigure}{0.45\textwidth}
        \includegraphics[width=\linewidth]{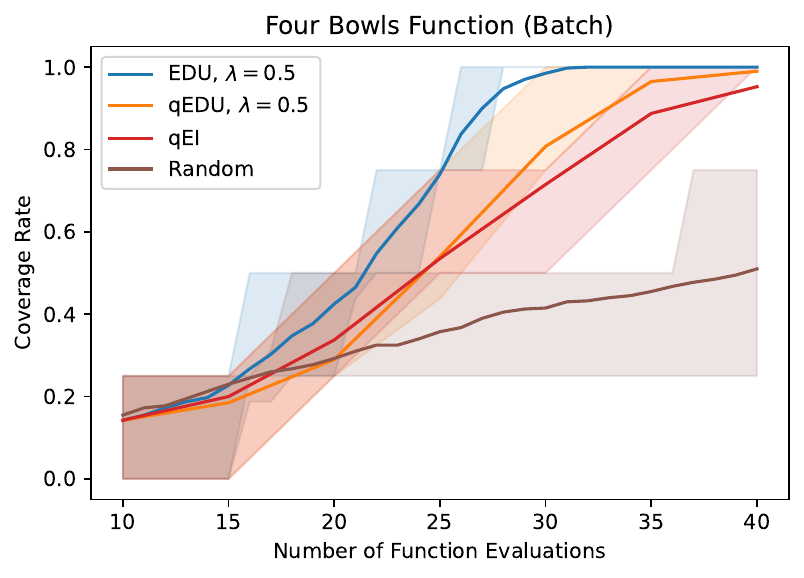}
    \end{subfigure}
    \hfill
    \begin{subfigure}{0.45\textwidth}
        \includegraphics[width=\linewidth]{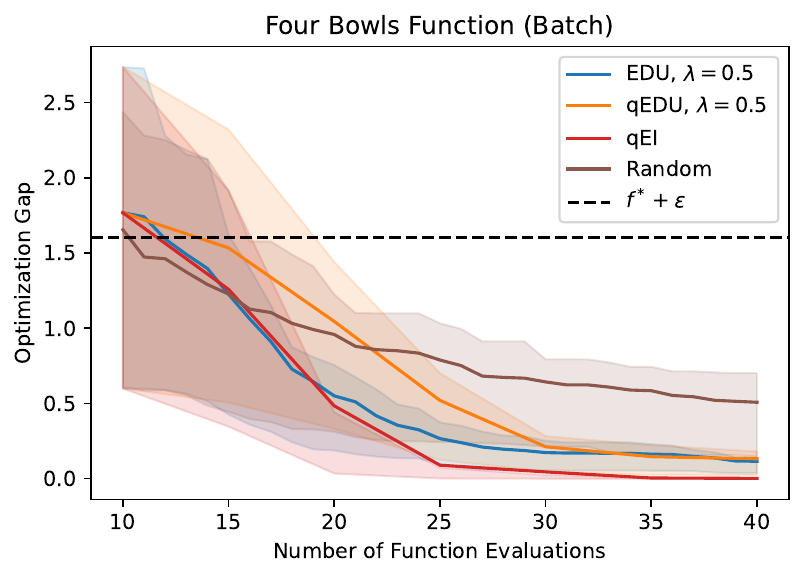}
    \end{subfigure}

    \vspace{1em} 

    \begin{subfigure}{0.45\textwidth}
        \includegraphics[width=\linewidth]{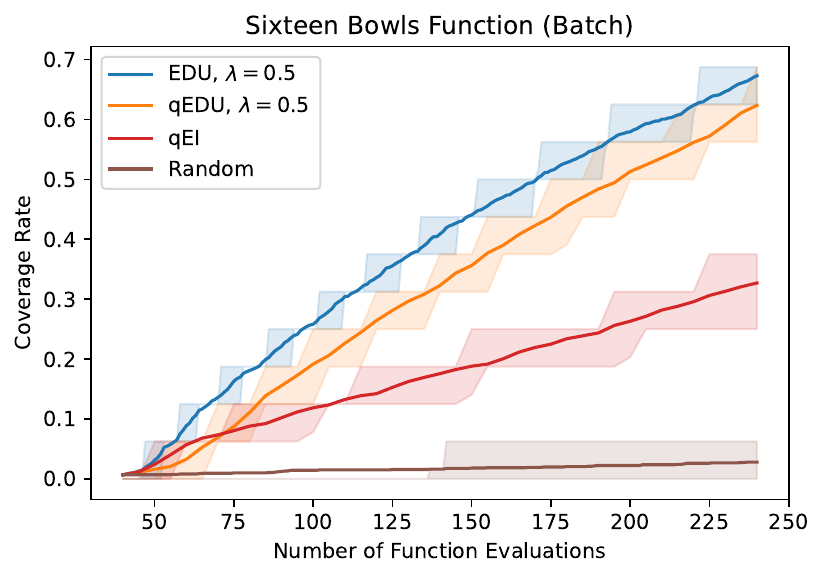}
    \end{subfigure}
    \hfill
    \begin{subfigure}{0.45\textwidth}
        \includegraphics[width=\linewidth]{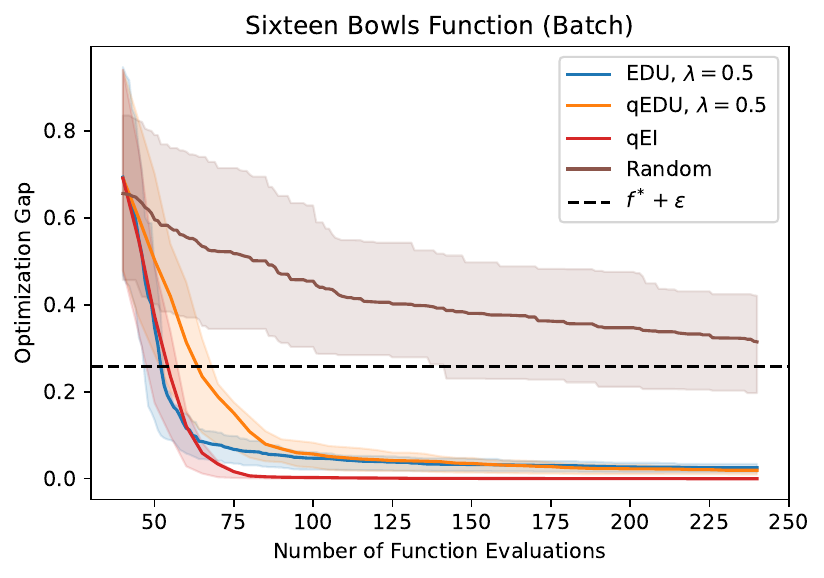}
    \end{subfigure}

    \caption{Solution coverage rates (left) and optimization gaps (right) of the EDU with $\lambda = 0.5$, $q$-EDU with $\lambda = 0.5$, $q$-EI and random sampling methods, for the $2^2$-bowl (top) and $2^4$-bowl (bottom) functions. A batch size of $q=5$ is used for the $q$-EI and $q$-EDU. Here, solid lines show the average metric over 100 simulations, with shaded regions showing the 25-th/75-th quantiles. The dotted lines on the right plots show the desired objective threshold $\gamma = f(\mathbf{x}^*) + \epsilon$.}
    \label{fig:bowls_performance_batch}
\end{figure}

Next, we investigate the performance of the batched $q$-EDU approach, compared with the batched $q$-EI from \cite{ginsbourger:hal-00260579}, both using a batch size of $q=5$. Figure \ref{fig:bowls_performance_batch} shows their performance in terms of solution coverage rates and optimization gaps. Aside from the first few sequential samples where $q$-EI yields slightly higher coverage rates as it hones in on the first optimal solution, the $q$-EDU yields considerably improved solution coverage and thus diverse optimization over the $q$-EI as more samples are collected. Both methods perform worse in terms of coverage rates compared to the (fully-sequential) EDU; this is not surprising, as the latter is provided the advantage of updating the GP surrogate fit after each sample point. Finally, the $q$-EDU yields significantly better diverse optimization performance over the baseline of random sampling, which is as desired.

\subsection{Six-Hump Camel Function}

Next, we explore a higher-dimensional test function with $d=8$ dimensions:
\begin{equation}
f(\mathbf{x}) = 2+\sum_{l=1}^4\left[\left(4-2.1\tau_l(\mathbf{x})^2 + \frac{\tau_l(\mathbf{x})^4}{3} \right)\tau_l(\mathbf{x})^2 + \tau_l(\mathbf{x})\eta_l(\mathbf{x}) + (-4+4\eta_l(\mathbf{x})^2)\eta_l(\mathbf{x})^2\right],
\label{eq:camel}
\end{equation}
where $\tau_l(\mathbf{x}) = x_{2(l-1)+1}$ and $\eta_l(\mathbf{x})=x_{2l}$. Here, each term summed in \eqref{eq:camel} is the ``six-hump camel'' test function \citep{surjanovic2013virtual}. Using a tolerance level of $\epsilon=f(\mathbf{x}^*)/10$, one can show that $f(\mathbf{x})$ has $K=16$ distinct $\epsilon$-optimal minima.

\begin{figure}[!t]
    \centering

    \begin{subfigure}{0.45\textwidth}
        \includegraphics[width=\linewidth]{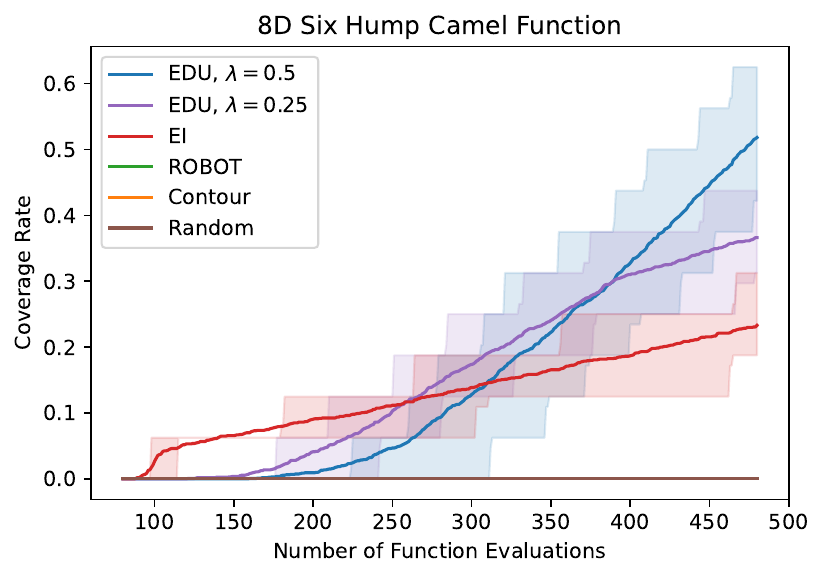}
    \end{subfigure}
    \hfill
    \begin{subfigure}{0.45\textwidth}
        \includegraphics[width=\linewidth]{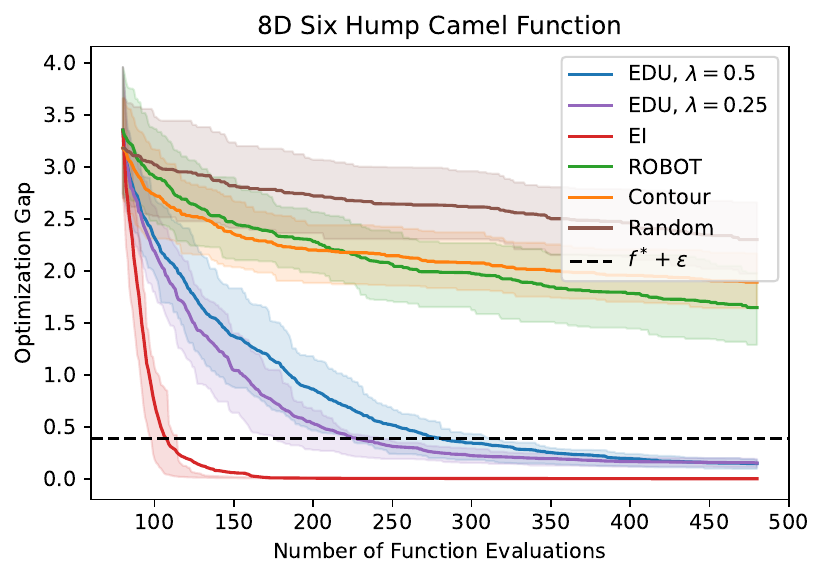}
    \end{subfigure}

    \caption{Solution coverage rates (left) and optimization gaps (right) of the EDU (with $\lambda = 0.5$ and $\lambda = 0.25$), EI, contour estimation, ROBOT and random sampling methods, for the $d=8$-dimensional six-hump camel function. Here, solid lines show the average metric over 100 simulations, with shaded regions showing the 25-th/75-th quantiles. The dotted line on the right plot shows the desired objective threshold $\gamma = f(\mathbf{x}^*) + \epsilon$. }
    \label{fig:6hump_performance}
\end{figure}

Figure \ref{fig:6hump_performance} shows the solution coverage rates and optimization gaps of the compared methods for this experiment. As before, the EDU achieves considerably better coverage rates (and thus improved diverse optimization) over the EI as more samples are collected. The EI again yields higher coverage initially as it hones in on the first optimal solution, but struggles in identifying other solutions. Here, the trade-off from $\lambda$ for the EDU (i.e., between diversity and pure optimization; see Section \ref{sec:acq}) is more evident. For larger $\lambda$ (i.e., $\lambda =0.5$), the EDU appears to place greater emphasis on solution diversity, which results in better solution coverage at the cost of a larger optimization gap. From our experiments, this trade-off from $\lambda$ appears to be more prominent as dimension increases, perhaps due to the increased potential for diverse solutions in higher dimensions. As before, ROBOT yields mediocre performance (for both solution coverage and optimization gap) with limited function evaluations.


\section{Applications}
\label{sec:app}
We now explore the effectiveness of the EDU in two applications, the first on rover trajectory optimization and the second on our motivating aviation engine control problem.

\subsection{Rover Trajectory Optimization}
Consider first the problem of path planning, which plays an important role in broad applications, including robotics, assembly planning and aircraft navigation \citep{path_planning_robotics,path_assembly,path_planning_flight}. Path planning aims to find optimal navigation paths through a terrain of various obstacles that inhibit movement. Such a problem is deceptively difficult; a key challenge is in evaluating the quality of a selected path, which typically requires costly computer simulations \citep{frank2008efficient}. Take, e.g., the navigation of an exploration rover on Mars \citep{carsten_rover}. Given a mapping of the terrain, a detailed simulation of a planned rover route can take minutes of computing time to reliably capture physical mechanisms \citep{frank2008efficient}. Furthermore, for path planning, it is highly preferable to obtain a ``basket'' of different viable routes; this facilitates path planning that is robust to unexpected changes in the operating environment or unforeseen obstacles.



\begin{figure}[!t]
    \centering
    \includegraphics[width = 7cm]{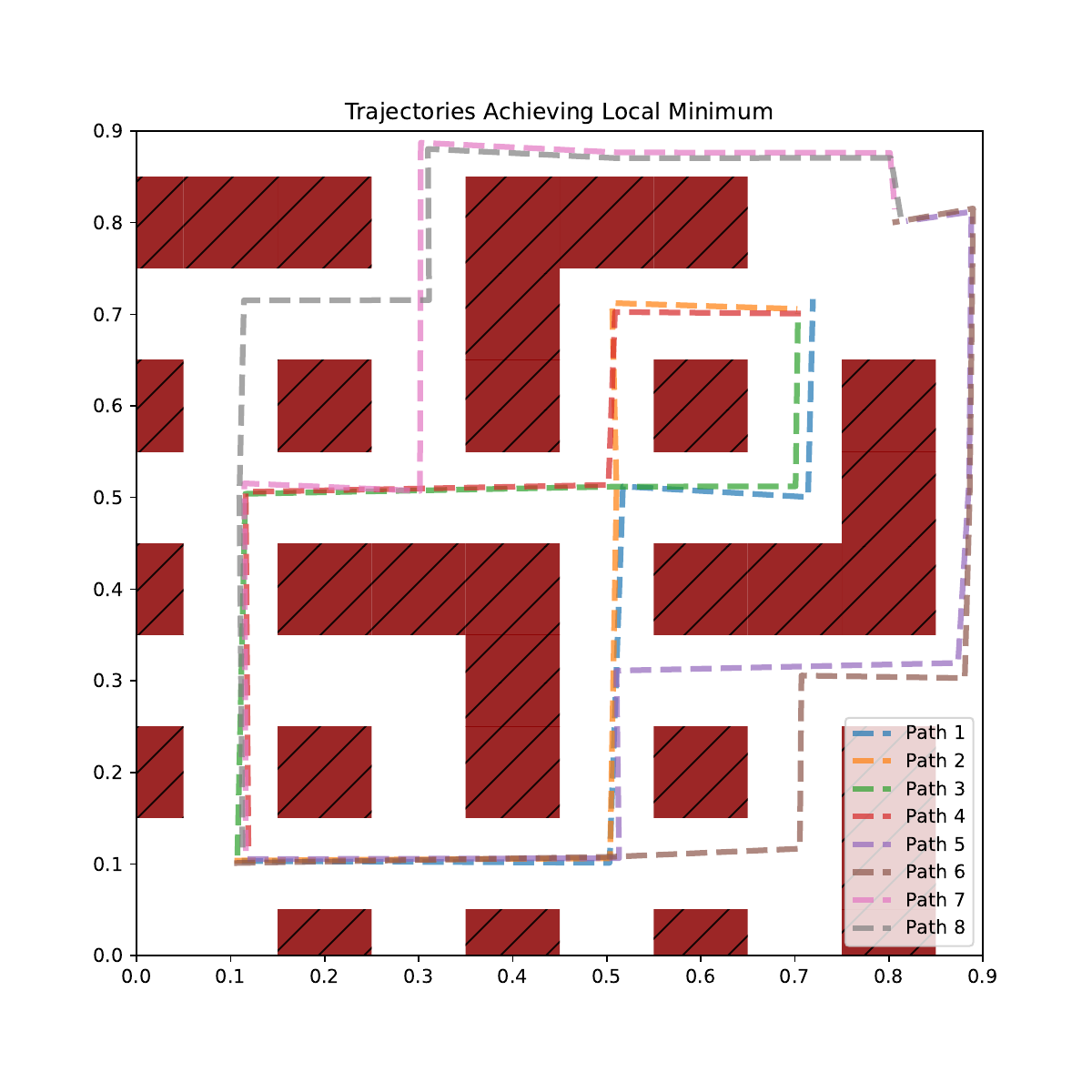}
    \caption{Visualizing the rover optimization problem. Here, red regions are obstacles that incur a penalty when hit. The $K=8$ $\epsilon$-optimal paths are shown in different colored dotted lines, with jitter for readability.}
    \label{fig:ROVTRAJ}
\end{figure}

We thus investigate the EDU in an adaptation of the rover trajectory optimization problem in \cite{wangrover,maus2023discovering}. The goal is to find a diverse basket of paths through the obstacle environment in Figure \ref{fig:ROVTRAJ}, taking the rover from the starting point $(0.05,0.05)$ to near the target destination point $(0.75, 0.75)$. The rover may pass through obstacles (colored in red), which incur a penalty when hit; more on this below. Simulation runs are performed using the Python path-planning module from \cite{wangrover}.

We consider here $d=12$ decision variables that parametrize the path trajectory, corresponding to six turn points for the rover; each turn point is parametrized by the distance travelled in the $x$- and $y$-directions. Such parameters are constrained within the range $[-1/15,1/3]$, where positive values indicate rightward and upward movement for $x$- and $y$-direction variables, respectively. This encourages turn angles that target rightward and upward movements to avoid inefficient path trajectories. The cost objective $f$ is then set as the scaled distance from the path's endpoint to the target coordinates of $(0.75,0.75)$, plus a penalty for hitting obstacles and minus a constant offset. This is given as follows:
\begin{equation}\label{eq:rover_cost_func}
f(\mathbf{x}) = 100\|\mathcal{P}_M(\mathbf{x})-(0.75,0.75)\|_2 + \frac{1}{2}\sum_{m=1}^{M-1}
\left(O\{ \mathcal{P}_m(\mathbf{x})\} + O\{\mathcal{P}_{m+1}(\mathbf{x}) \}\right)\|\mathcal{P}_m(\mathbf{x})-\mathcal{P}_{m+1}(\mathbf{x})\|_2 - 5.
\end{equation}
Here, the considered path (parametrized by variables $\mathbf{x}$) is discretized into $M$ timesteps, with $\mathcal{P}_m(\mathbf{x})$ denoting the path coordinates at timestep $m$. For the penalty term, $O\{ \mathcal{P}_m(\mathbf{x}) \} = 30 \times \mathbbold{1}\{\text{$\mathcal{P}_m(\mathbf{x})$ is on an obstacle}\}$ + 0.05; $f(\cdot)$ is thus penalized at a rate of 30 if the path is on an obstacle. Finally, the error tolerance is set as $\epsilon=17$; this allows us to find trajectories that are close to the destination while passing through at most one or two obstacles. With this $\epsilon$, Figure \ref{fig:ROVTRAJ} shows the $K=8$ $\epsilon$-optimal paths. These paths are relatively distinct from one another, and the goal is to identify such paths using a limited number of simulation runs.

As before, we compare the performance of the EDU (with default setting $\lambda = 0.5$) with the earlier four baseline methods: the EI, ROBOT, contour estimation and random sampling, all with the aforementioned recommended settings. Each method starts with an initial Latin hypercube design of $n=60$ points, then proceeds with 720 additional sequential runs. This experiment is then replicated 100 times to provide simulation variability. For this problem, note that the set of $K=8$ $\epsilon$-optimal solutions is known a priori (Figure \ref{fig:ROVTRAJ}), thus our comparison metrics, i.e., solution coverage rate and optimization gap, can be evaluated exactly. We can further determine the tolerability of an evaluated path $\mathbf{x}$, since the global minimum $f(\mathbf{x}^*)$ is known. To foreshadow, such comparisons become more challenging in our later motivating flight application, where the true solutions are not know beforehand.






\begin{figure}[!t]
    \centering

    \begin{subfigure}{0.49\textwidth}
        \includegraphics[width=\linewidth]{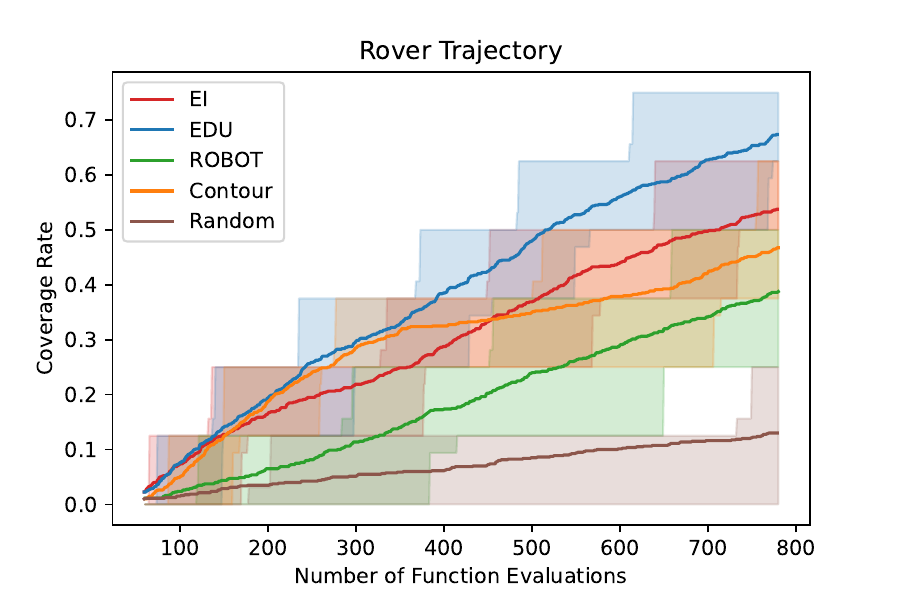}
    \end{subfigure}
    \hfill
    \begin{subfigure}{0.49\textwidth}
        \includegraphics[width=\linewidth]{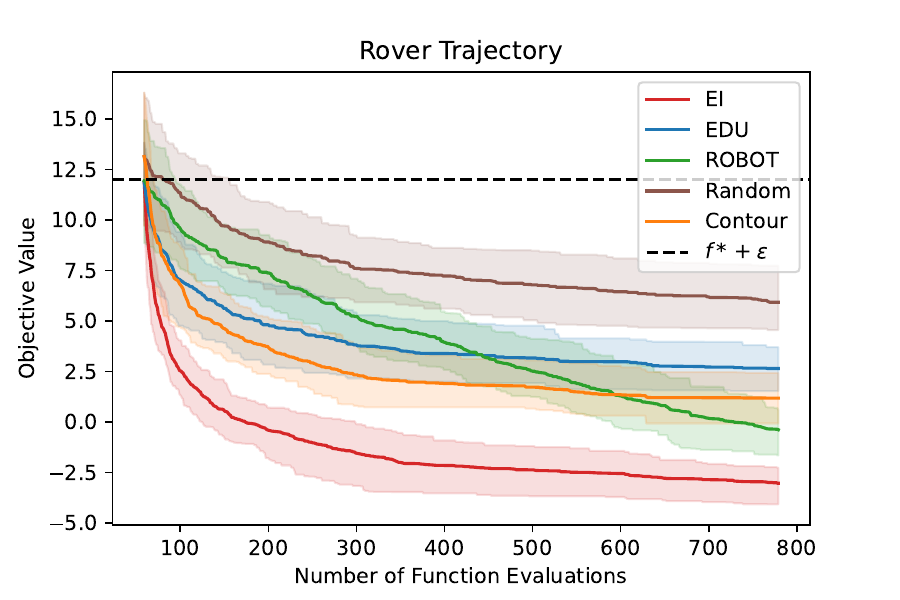}
    \end{subfigure}
    \caption{Solution coverage rates (left) and optimization gaps (right) of the EDU, EI, contour estimation, ROBOT and random sampling methods, for the rover trajectory application. Here, solid lines show the average metric over 100 simulations, with shaded regions showing the 25-th/75-th quantiles. The dotted line on the right plot shows the desired objective threshold $\gamma = f(\mathbf{x}^*) + \epsilon$. }
    \label{fig:ROVRES}
\end{figure}

Figure \ref{fig:ROVRES} shows the solution coverage rates (for the $K=8$ $\epsilon$-optimal paths in Figure \ref{fig:ROVTRAJ}) and optimization gaps for this rover trajectory application. As before, we see that the EDU offers much improved diverse optimization performance (i.e., higher coverage rates) compared to existing methods, which is desirable. This improvement comes at the expense of global optimization performance, but only after the desired objective tolerance (dotted black line) is achieved. ROBOT and random sampling again yields mediocre diverse optimization performance (i.e., smaller coverage rates); the former is not too surprising given that it requires rather large sample sizes to achieve diversity. Here, the contour method yields comparatively better diverse optimization compared to earlier simulations, but is still outperformed by both the EI and EDU. 


\begin{figure}[!t]
    \centering
    \includegraphics[width = 15cm]{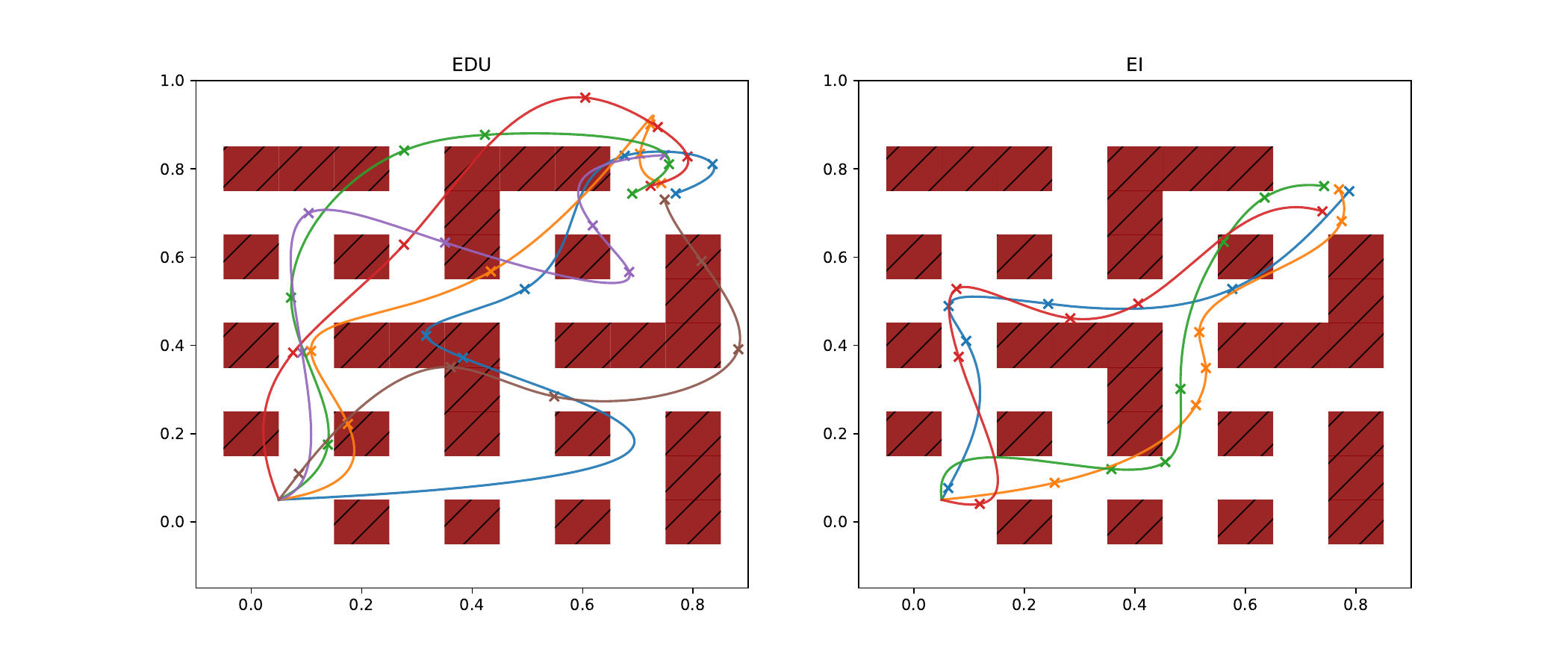}
    
    \caption{Visualizing representative tolerable rover paths found by the EDU (left) and EI (right). Here, turn points within a path are marked by an \texttt{x}.}

    \label{fig:ROVER_EI_DEI_COMPARISON}
\end{figure}

To better visualize the improved path diversity from EDU, Figure \ref{fig:ROVER_EI_DEI_COMPARISON} shows several representative tolerable paths found by the EDU and EI. Here, the EI identifies paths that largely avoid any obstacles and end close to the destination point $(0.75,0.75)$, which is in line with its excellent global optimization performance noted earlier. However, it fails to identify the diverse collection of possible paths desired from Figure \ref{fig:ROVTRAJ}. The EDU, in sacrificing some global optimality performance, is able to identify a larger ``basket'' of diverse paths; this supports the excellent diverse optimization performance noted earlier. Thus, given limited simulation runs, the EDU provides a promising approach for diverse optimization compared to existing black-box methods.

\subsection{Diverse Optimization for Internal Combustion Engine Control}
\label{sec:ice}

We now tackle our motivating application on engine control for sustainable aviation. Recall from Section \ref{sec:mot} that there are $d=4$ control parameters for engine control: start-of-injection ($x_1$, SOI), glow plug power ($x_2$, GPP), engine speed ($x_3$, in RPMs) and fuel cetane number ($x_4$, CN). Our need for diverse solutions arises from a myriad of restrictions on fuel blends and energy consumption (and thus on $x_1-x_4$) that may change during operation. Here, diverse optimization offers a promising solution: for a carefully-specified tolerance level $\epsilon>0$, the basket of solutions $\mathbf{x}_1^*, \cdots, \mathbf{x}_K^*$ provides optimal control strategies over different tolerable regions, thus facilitating timely \textit{accommodation} of changing constraints in real-time and \textit{planning} of flight strategies for fuel efficiency.


To do so, we first require a measure for flight instability, $f(\mathbf{x})$, which we aim to minimize with respect to a control strategy $\mathbf{x}$. Recall from Section \ref{sec:prob} that the considered engine system cycle can be measured from -360 to +360 crank-angle-degrees (CADs), where 0 CAD refers to the top-dead-center position where the engine piston is in its most compressed state. A natural instability metric from past studies \citep{sapra2023numerical,narayanan2024misfire,pal2024data} leverages the so-called CA50, the CAD at which 50\% of the fuel's energy has been released. The CA50 is a widely-used metric for engine performance, particularly within the compression ignition engine community for controller calibration \citep{dong2022data,govindrajurate2023}. We thus target the diverse minimization of this CA50-based metric $f$, which is a direct output of the virtual simulator described in Section \ref{sec:prob}. Here, the tolerance level $\epsilon$ is set as 0.9 CAD, following the recommended acceptable error tolerance suggested in \cite{dong2022data} for flight control.


For initial design, all methods begin with the same $n=40$ points, generated from a maximum projection design \citep{joseph_maxpro}. Sequential samples are then generated from each method in batches of $q=5$, which was set from our HPC allocation; 6 batches were generated, yielding a total of 30 sequential samples. Since each simulation run is highly cost-intensive (requiring over 3,000 CPU hours), we compare the $q$-EDU (the batched version of EDU) with only the $q$-EI (the batched version of EI), which performed best out of the existing methods in earlier simulations. As before, the $q$-EDU is performed using the default setting of $\lambda = 0.5$.

Here, we face two challenges for methods comparison. First, unlike in earlier experiments, the set of true $\epsilon$-optimal solutions $\mathbf{x}_1, \cdots, \mathbf{x}_K$ are unknown here, hence the comparison metrics, i.e., coverage rate and optimization gap, cannot be directly evaluated. We thus investigate below alternate metrics for comparison, guided by the target needs of our flight control application. Second, since the true global minimum $f(\mathbf{x}^*)$ is not known here, another challenge is how to determine the tolerability of an evaluated solution $\mathbf{x}$. We adopt the lower bound approach outlined in Section \ref{sec:eps}. Here, it is known \citep{dong2022data} that the flight instability metric $f(\mathbf{x})$ is non-negative, and that an ideal control strategy $\mathbf{x}^*$ should have a near-zero instability metric of $f(\mathbf{x}^*)$. For the below analysis, we thus deem an evaluated solution $\mathbf{x}$ tolerable if $f(\mathbf{x}) \leq f_L + \epsilon$, where $f_L = 0$ is the elicited lower bound.

\begin{figure}[!t]
    \centering

    \begin{subfigure}{0.33\textwidth}
        \includegraphics[width=\linewidth]{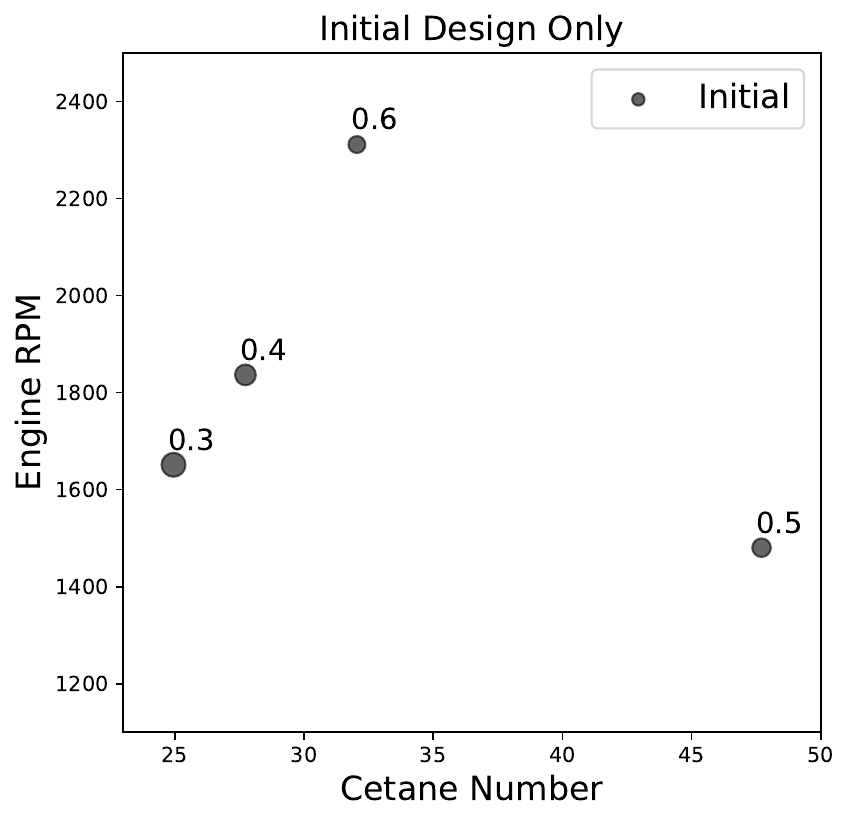}
    \end{subfigure}
    \hfill
    \begin{subfigure}{0.33\textwidth}
        \includegraphics[width=\linewidth]{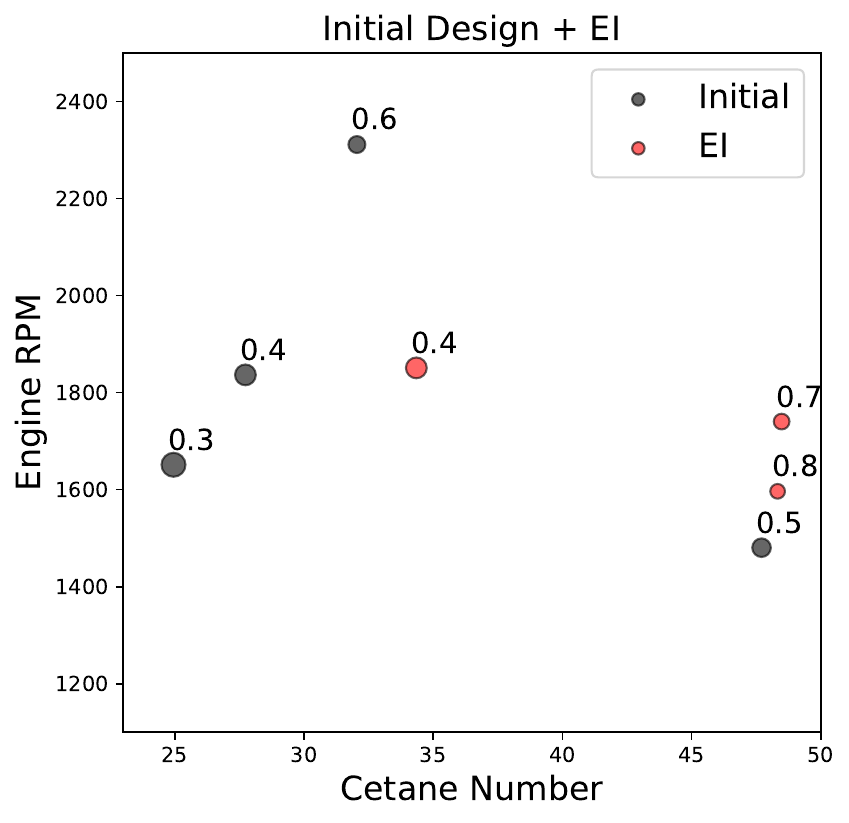}
    \end{subfigure}
    \hfill
    \begin{subfigure}{0.33\textwidth}
        \includegraphics[width=\linewidth]{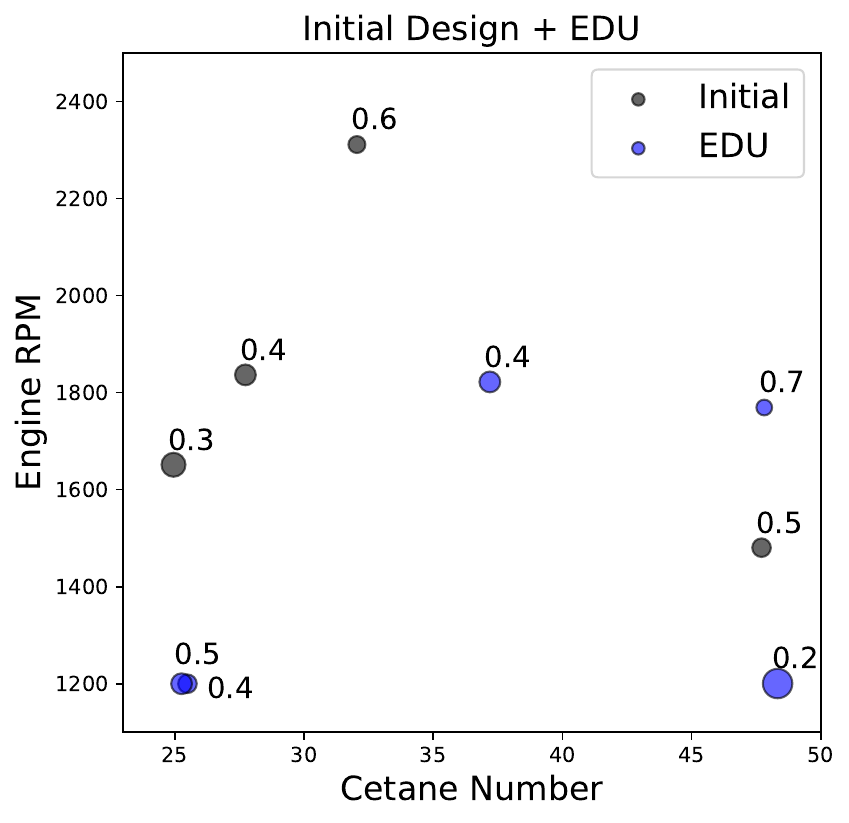}
    \end{subfigure}
    
    \caption{Visualizing the identified tolerable solutions projected onto $(x_3,x_4)$-space from initial design points (left), initial with EI points (middle), and initial with EDU points (right). Here, a larger point size indicates a smaller (i.e., better) objective value for $f$; this value is provided next to the point.}
    \label{fig:CFD_scatter}
\end{figure}

We first inspect the tolerable control solutions $\mathbf{x}$ found by each method. It is known \citep{pal2024data,narayanan2024simulation,govindrajurate2023} that the changing restrictions on fuel blends and energy consumption during flight (see Section \ref{sec:need}) most affect two specific control parameters: engine speed ($x_3$) and fuel cetane number ($x_4$). As such, it is critical that the identified control solutions cover well this $(x_3,x_4)$-space, so that a good control strategy is available given changing restrictions on such a space during flight. Figure \ref{fig:CFD_scatter} shows the identified tolerable solutions from $q$-EDU and $q$-EI; which we refer to as EDU and EI hereafter for brevity. These solutions are projected onto the desired $(x_3,x_4)$-space, with larger point sizes reflecting smaller objectives for $f(\mathbf{x})$, i.e., better control solutions.



Consider first the \textit{coverage} of the identified solutions over the $(x_3,x_4)$-space in Figure \ref{fig:CFD_scatter}. We see that, within the initial $n=40$ runs, there are four tolerable control solutions. For the subsequent 30 batch-sequential runs, the EI identifies three additional tolerable solutions (totaling seven), while the EDU identifies five additional tolerable solutions (totaling nine). From Figure \ref{fig:CFD_scatter}, we further see that the nine identified solutions from the EDU provide better coverage over the $(x_3,x_4)$-space: it pinpoints tolerable control solutions on the bottom-left and bottom-right corners of this 2d-space, whereas the EI fails to find any solutions there. Having viable control strategies within these two corners is critical for stable flight performance. The bottom-left corner corresponds to a scenario with low RPMs (i.e., low $x_3$) and low fuel cetane numbers (i.e., low $x_4$). Here, low RPMs reflect typical flight cruise conditions that make up a large part of flight operation, and low cetane numbers indicate high concentrations of sustainable aviation fuels. Such a scenario poses a significant challenge for flight due to the aforementioned ignition difficulties for such sustainable fuels \citep{stafford2023combined,miganakallu2022impact}. By finding a viable control strategy in this corner, the EDU addresses this critical flight bottleneck with sustainable fuels; the EI, on the other hand, fails to find a tolerable solution, which increases the risk of disastrous engine misfires. The bottom-right corner corresponds to a scenario with low RPMs and high cetane numbers. Here, high cetane numbers indicate the use of high concentrations of diesel-based fuels. Such high cetane fuels are needed when energy-intensive glow plugs are infeasible to use for energy assistance, e.g., during periods of strict energy demands or glow plug overuse. Again, the EDU finds a viable control strategy here, whereas the EI fails to find a tolerable solution, which may lead to real-time flight instabilities. In this sense, the EDU offers a more complete basket of strategies for real-time control and flight planning to ensure flight stability.




This improved coverage of the EDU can be numerically assessed via two ``space-fillingness'' metrics on the identified solutions in Figure \ref{fig:CFD_scatter}. For a given method, let $\mathcal{T}$ denote its identified tolerable solutions. Define these two metrics as:
\begin{equation}
\text{SF}_1(\mathcal{T}) := \max_{\mathbf{x} \in [0,1]^d} Q(\mathbf{x};\mathcal{T}), \quad \text{SF}_2(\mathcal{T}) := \int_{[0,1]^d} Q(\mathbf{x};\mathcal{T}) \; d\mathbf{x},
\label{eq:space}
\end{equation}
\noindent where $Q(\mathbf{x};\mathcal{T}) = \min_{\mathbf{x}' \in \mathcal{T}} \|\mathbf{x}-\mathbf{x}'\|_2$ denotes the distance between $\mathbf{x}$ and its closest point in $\mathcal{T}$. In words, $\text{SF}_1(\mathcal{T})$ measures the distance of the farthest point on the parameter space $[0,1]^d$ to $\mathcal{T}$, and $\text{SF}_2(\mathcal{T})$ measures its integrated analogue. Thus, the smaller these metrics are, the better $\mathcal{T}$ covers the desired space; see \cite{mak2018minimax} for details. Such metrics extend naturally for assessing coverage on projections of $[0,1]^d$. Table \ref{tab:CFD_Results} reports these metrics over the full $[0,1]^4$ space as well as its projection onto $(x_3,x_4)$-space. We see that the EDU solutions indeed yield noticeably better coverage metrics than the EI, both on the full space and the $(x_3,x_4)$-space. This is expected from Figure \ref{fig:CFD_scatter}, and suggests the EDU offers a more diverse basket of strategies for real-time flight control.

Consider next the \textit{quality} of the identified solutions for minimizing $f$. This is visualized in Figure \ref{fig:CFD_scatter}, where a larger point size indicates a better control solution with smaller objective $f$. Here, we see that the plotted EDU solution points are noticeably larger (and thus have smaller objective values) than the EI. This suggests that, in addition to identifying more diverse solutions with better \textit{coverage} of the control space, the EDU further provides higher \textit{quality} control solutions that better minimize flight instability (as measured by $f$). This is quite surprising given earlier experiments, where the EI outperformed the EDU in terms of global optimization performance. In retrospect, a plausible reason for this is that the response surface $f$ appears to have multiple near-optimal control solutions along the boundaries of the input space. For example, the best solution found by EDU is $\mathbf{x}=(-1.68,70,48,1200)$, which is a boundary point. As the EDU places greater emphasis on diversity, i.e., exploration within the $\epsilon$-optimal region, this may have encouraged greater exploration of boundary regions, thus leading to better global optimization performance.

\begin{table}[!t]
        \caption{
        Summarizing the two space-filling metrics $\text{SF}_1$ and $\text{SF}_2$ from \eqref{eq:space} for the initial design points, initial with EI points, and initial with EDU points. These metrics are provided on the projected $(x_3,x_4)$-space and on the full four-dimensional space.
        }
    \centering
    \begin{tabular}{lcccc}
        \toprule
         & \multicolumn{2}{c}{$(x_3,x_4)$-space} & \multicolumn{2}{c}{Full 4-d space} \\
        \cmidrule(lr){2-3} \cmidrule(lr){4-5}
         & $\text{SF}_1$ & $\text{SF}_2$ & $\text{SF}_1$ & $\text{SF}_2$ \\
        \midrule
        Initial Design & 0.6929 & 0.2967 & 1.1467 & 0.5666 \\
        Initial Design + $q$-EI & 0.5648 & 0.2343 & 1.1467 & 0.5074  \\
        \textbf{Initial Design + $q$-EDU} & \textbf{0.5463} & \textbf{0.2122} & \textbf{0.9981} & \textbf{0.4796}  \\
        \bottomrule
    
    \end{tabular}

    \label{tab:CFD_Results}
\end{table}

\section{Conclusion}
\label{sec:conc}

We propose a novel method, called Expected Diverse Utility, which tackles the timely problem of diverse black-box optimization of expensive computer simulators. The EDU addresses an urgent need for diverse control strategies in sustainable aviation, where an operator faces changing constraints on control parameters during flight. The EDU builds upon the popular Expected Improvement approach to search for diverse ``$\epsilon$-optimal'' solutions -- locally optimal solutions within a tolerance $\epsilon > 0$ from the global optimum. Under a Gaussian process surrogate model, the EDU can be evaluated analytically in closed-form, thus allowing for efficient acquisition optimization for subsequent evaluation points via automatic differentiation. This closed-form expression further reveals a novel exploration-exploitation-diversity trade-off, which extends the well-known exploration-exploitation trade-off in reinforcement learning. We then show the improved diverse optimization performance of the EDU over existing methods in a comprehensive suite of numerical experiments. Finally, we demonstrate the effectiveness of the EDU in two applications, the first on rover trajectory optimization and the second on our motivating application for real-time flight control.

Such promising developments open the door for exciting opportunities in methods development and applications in broad scientific and engineering problems. There has been much recent work on improved surrogate modeling via multi-fidelity simulators \citep{ji2022multi} and manifold learning \citep{zhang2022gaussian,li2023additive}; we aim to leverage such developments for improving the EDU in high-dimensional diverse optimization. In addition to aviation control, we are exploring broader applications of the EDU for materials optimization and policy planning. Diverse black-box optimization can play an important role in guiding timely decision-making in these fields, e.g., it can help identify a ``basket'' of promising material designs or policy decisions for downstream experimentation or implementation. We defer this to future work.\\

\noindent \textbf{Acknowledgements.} SM and JM are supported by NSF CSSI 2004571, NSF DMS 2210729, NSF DMS 2220496, NSF DMS 2316012 and DE-SC0024477. SRN, SY and ZS are supported by the Army Research Laboratory (ARL) under Cooperative Agreement Number W911NF2020161. The views and conclusions contained in this document are those of the authors and should not be interpreted as representing the official policies, either expressed or implied, of the Army Research Laboratory or the U.S. Government. The U.S. Government is authorized to reproduce and distribute reprints for Government purposes notwithstanding any copyright notation herein. SRN and SY would like to thank Dr. Harsh Sapra, Dr. Randy Hessel and Dr. Sage Kokjohn from the University of Wisconsin-Madison for their valuable assistance with the CFD case setup. 
SRN and SY would finally like to thank Convergent Science for providing CONVERGE 3.0 licenses and technical support for the CFD simulations used in this work.

\bibliography{DEI_references}       

\end{document}